\pdfoutput=1
\documentclass[aps,prd,a4paper,showpacs,showkeys,superscriptaddress, preprintnumbers,floatfix,nofootinbib,twocolumn]{revtex4-1}

\usepackage[toc,page]{appendix}
\usepackage{latexsym}
\usepackage{revsymb}
\usepackage{multirow}
\usepackage{color}
\usepackage[usenames,dvipsnames,svgnames,table]{xcolor}

\usepackage{graphicx}
\usepackage{epsfig}  
\usepackage{epsf}    
\usepackage{dcolumn}
\usepackage{bm}
\usepackage{dcolumn}
\usepackage{textcomp}
\usepackage{float}
\usepackage[]{hyperref}
\usepackage{adjustbox}
\usepackage[utf8]{inputenc}
\usepackage[T1]{fontenc} 
\usepackage{url}
\usepackage{amsmath}
\usepackage{amssymb}
\usepackage{mathtools}
\usepackage{subfigure}
\usepackage{alphalph}
\usepackage{morefloats}
\usepackage{comment}
\usepackage{ulem}

\colorlet{shadecolor}{gray!15}
\definecolor{greenLinks}{rgb}{0, 0.6, 0} 
\definecolor{blueLinks}{rgb}{0, 0, 0.6}
\definecolor{redLinks}{rgb}{0.6, 0, 0}
\definecolor{tempText}{rgb}{0.55, 0.10,0.67}
\definecolor{eprintLinks}{rgb}{0.4, 0.4, 0.4}
\definecolor{journalLinks}{rgb}{0.6, 0, 0}

\usepackage[T1]{fontenc}

\def\mg#1{\textcolor{magenta}{#1}}

\usepackage{hyperref}
\hypersetup{colorlinks=true,linkcolor=redLinks,citecolor=greenLinks,
urlcolor=redLinks,
pdfborder={0 0 1}}
\hypersetup{
    bookmarks=true,         
    unicode=false,          
    pdftoolbar=true,        
    pdfmenubar=true,        
    pdffitwindow=false,     
    pdfstartview={FitH},    
    pdfauthor={Author},     
    pdfsubject={Subject},   
    pdfcreator={Creator},   
    pdfproducer={Producer}, 
    pdfkeywords={keyword1, key2, key3}, 
    pdfnewwindow=true,      
    colorlinks=false,       
    linkcolor=red,          
    citecolor=green,        
    filecolor=magenta,      
    urlcolor=cyan           
}

\def\SM{$\mathrm{SU(3)_c \otimes SU(2)_L \otimes U(1)_Y}$ }
\def\21{$\mathrm{SU(2)_L \otimes U(1)_Y}$ }

\def\lnv{lepton number violating }

\newcommand{\AddrAHEP}{  AHEP Group, Institut de F\'{i}sica Corpuscular --
  CSIC/Universitat de Val\`{e}ncia, Parc Cient\'ific de Paterna.\\
 C/ Catedr\'atico Jos\'e Beltr\'an, 2 E-46980 Paterna (Valencia) - Spain}
\newcommand{\AddrMiranda}{%
Departamento de F\'{\i}sica, Centro de Investigaci\'on
  y de Estudios Avanzados del IPN,\\ Apartado Postal 14-740 07000 Mexico,
  Distrito Federal, Mexico}
  
\newcommand{\AddrSabyaIPhT}{Institut de Physique Th{\'e}orique, Universit{\'e} Paris Saclay, CNRS, CEA, F-91191 Gif-sur-Yvette, France}  
\newcommand{\AddrSabyaIPPP}{Institute for Particle Physics Phenomenology, Department of Physics, Durham University, Durham, DH1 3LE, UK}

\newcommand{\AddrFisteo}{%
Departament de F\'{i}sica Te\'orica, Universitat de Val\`encia, Burjassot 46100, Spain}

\begin{document}

\preprint{t21/073}

\title{Nonunitarity of the lepton mixing matrix at the European Spallation Source}

\author{Sabya Sachi Chatterjee} \email{sabya-sachi.chatterjee@ipht.fr} \affiliation{\AddrSabyaIPhT}\affiliation{\AddrSabyaIPPP}

\author{O. G. Miranda} \email{omr@fis.cinvestav.mx} \affiliation{\AddrMiranda}

\author{M. T\'ortola} \email{mariam@ific.uv.es} \affiliation{\AddrFisteo}\affiliation{\AddrAHEP}
\author{J. W. F. Valle}\email{valle@ific.uv.es} \affiliation{\AddrAHEP}
%

\begin{abstract}
  If neutrinos get mass through the exchange of lepton mediators, as in seesaw schemes, the neutrino appearance probabilities in oscillation experiments are modified due to effective nonunitarity of the lepton mixing matrix.
  This also leads to new CP phases and an ambiguity in underpinning the ``conventional'' phase of the three-neutrino paradigm. We study the CP sensitivities of various setups based at the European spallation source neutrino super-beam (ESSnuSB) experiment in the presence of nonunitarity.
  We also examine its potential in constraining the associated new physics parameters. Moreover, we show how the combination of DUNE and ESSnuSB can help further improve the
    sensitivities on the nonunitarity parameters.
\end{abstract}


\maketitle
\section{Introduction}
\label{sec:intro}
The discovery of neutrino oscillations~\cite{McDonald:2016ixn,Kajita:2016cak} has brought neutrinos to the center of particle physics.
The current experimental data mainly converge into a consistent global picture in which the oscillation parameters are pretty well determined.
However, three challenges still remain, namely, to determine the CP phase, the atmospheric octant and the ordering of the neutrino mass spectrum~\cite{deSalas:2020pgw,10.5281/zenodo.4726908}.
These will be the target of a number of future experiments, such as DUNE~\cite{DUNE:2015lol}.
A fourth item must be added to this list, namely probing the robustness of the interpretation, such as testing the unitarity of the lepton mixing matrix. 
This is crucial because it undermines the efforts of underpining the CP phase $\delta_{\rm CP}$~\cite{Miranda:2016wdr,Escrihuela:2016ube}. 

This task is well justified also on theory grounds.  Indeed, one of the most attractive ways to generate neutrino mass is through the mediation of heavy neutral leptons. 
While these emerge in many gauge extensions of the standard model, they can be postulated directly at the \SM\ level, as the neutrino mass generation mediators. 

This, in fact, provides the most general realization of the seesaw mechanism and many of its variants~\cite{Schechter:1980gr}. 
For generality here we focus exclusively on this case, namely, the standard \SM\ seesaw mechanism. 
The resulting lepton mixing matrix is in general quite complex when compared with Cabibbo-Kobayashi-Maskawa (CKM) mixing. 
First, lepton mixing contains extra phases that can not be eliminated by field redefinitions~\cite{Schechter:1980gr} and are therefore physical~\cite{Schechter:1980gk},
crucially affecting \lnv processes. However, they do not affect conventional oscillations, so we will ignore them in what follows.
On the other hand the lepton mixing matrix must in general take into account the admixture of the heavy lepton seesaw mediators with the light active neutrinos~\cite{Schechter:1981cv}. 
These are usually neglected, as the smallness of neutrino masses indicated by neutrino experiments suggests a very high seesaw scale.  

Nonetheless, the seesaw mechanism can also be realized at low scales. 
The template for this is a scenario where two SM-singlet leptons are added sequentially, instead of just one.
If lepton number symmetry is imposed, then all three active neutrinos are massless, as in the standard model.  
In contrast to the Standard Model, however, lepton flavor is violated, and similarly, leptonic CP symmetry.  
This shows that flavor and CP violation can exist in the leptonic weak interaction despite the masslessness of neutrinos, implying that such processes
need not be suppressed by the small neutrino masses, and hence can be large~\cite{Bernabeu:1987gr,Branco:1989bn,Rius:1989gk,Dittmar:1989yg,Deppisch:2004fa,Deppisch:2005zm}.

Over such basic template one can build genuine ``low-scale'' realizations of the seesaw mechanism
in which lepton number symmetry is restored at low, instead of high, values of the \lnv scale.  
The models are natural in t'Hooft sense, and lead to small, symmetry-protected neutrino masses. 
Such ``low-scale'' seesaw realizations include the inverse~\cite{Mohapatra:1986bd,GonzalezGarcia:1988rw}
as well as the linear seesaw mechanisms~\cite{Akhmedov:1995ip,Akhmedov:1995vm,Malinsky:2005bi}.
In all of these we expect potentially sizeable unitarity violation in the leptonic weak interaction. 
This paper is dedicated to probing such effects at the European Spallation Source neutrino Super-Beam (ESSnuSB) experiment.
As far as we can tell, this is the first study of this kind for ESSnuSB.
In addition, we also \mg{review} the sensitivities of the DUNE experiment to the nonunitarity parameters and show how DUNE and ESSnuSB can play a complementary role to each other.
Sensitivity studies of nonunitarity at other future neutrino facilities can be found in Refs.~\cite{Escrihuela:2016ube,Ge:2016xya,Meloni:2009cg, Blennow:2016jkn,Miranda:2018yym,Soumya_2018,Miranda:2020syh,Coloma:2021uhq,Agarwalla:2021owd}.

We briefly describe the theoretical framework for unitarity violation in the charged current (CC) leptonic weak interaction in Sec.~\ref{sec:theor-fram},
and the matter three-neutrino oscilation probabilities with and without unitarity violation in Sect.~\ref{sec:probability}.
Next, in Sec.~\ref{sec:experimental-setup}, we describe the experimental setups of interest, and also present the details of the simulation we have performed.
Our results are given in Sec.~\ref{sec:results} and include our calculated ESSnuSB and DUNE sensitivities to nonunitary (NU) neutrino mixing in Sec.~\ref{sec:essn-sens-non}, 
the CP violation discovery potential in the presence of unitarity violation is given in~\ref{sec:cp-viol-disc}, and the CP reconstruction capabilities both for the standard phase as well as the seesaw phase 
of $\alpha_{21}$ in~\ref{sec:cp-reconstruction}. Finally we briefly summarize in Sec.~\ref{sec:Conclusions}.

\section{Theoretical framework}
\label{sec:theor-fram}
In the standard $3\times3$ oscillation picture, the neutrino mixing matrix is described symmetrically by a product of three mixing matrices 
\begin{equation}
U^{}=\omega_{2\,3}\:\omega_{1\,3}\:\omega_{1\,2} \, ,
\end{equation}
where each $\omega_{i,j}$ describes an effective $2\times 2$ complex rotation, characterized by a mixing angle and its phase. This symmetrical form complements the original description~\cite{Schechter:1980gr}
by specifying the most convenient factor ordering. In explicit form, the standard $3\times3$ leptonic mixing matrix is given by
\begin{widetext}
\begin{equation}
   \label{eq:rot}
U =\left(\begin{array}{ccccc}
  1   & 0 & 0 \\
  0   & c_{23} & e^{-i \varphi_{23}}s_{23}   \\
  0   & -e^{i \varphi_{23}}s_{23} & c_{23}
\end{array}\right) 
\left(\begin{array}{ccccc}
c_{13} & 0 & e^{-i \varphi_{13}}s_{13}   \\
  0   & 1 & 0                    \\
-e^{i \varphi_{13}} s_{13} & 0 & c_{13}
\end{array}\right) 
\left(\begin{array}{ccccc}
c_{12} & e^{-i \varphi_{12}}s_{12} &  \\
-e^{i \varphi_{12}} s_{12} & c_{12} & 0  \\
  0   & 0 & 1       
\end{array}\right) ,
\end{equation}
\end{widetext}
where $s_{ij}=\sin\theta_{ij}$, $c_{ij}=\cos\theta_{ij}$, and $\varphi_{ij}$ is the corresponding phase. 
One sees the appearance of two extra physical phases with no counterpart in the quark sector: the so-called Majorana phases~\cite{Schechter:1980gr}.
Note that the above parameterization of the neutrino mixing matrix is equivalent to the oscillation-sensitive part of the PDG form~\cite{Zyla:2020zbs}
  with $\varphi_{13} - \varphi_{12} - \varphi_{23} \equiv \delta_\mathrm{CP}$~\cite{Rodejohann:2011vc} so that when $\varphi_{12} = \varphi_{23} = 0$ one has $\varphi_{13} = \delta_\mathrm{CP}$.

Apart from the presence of these new physical~\cite{Schechter:1980gk} phases, the leptonic CC interaction will in general also contain the mixing of neutral heavy leptons that mediate neutrino mass generation, as in the so-called type-I seesaw mechanism.
These two facts make the mixing of massive neutrinos substantially richer in structure than that which describes the quark weak interactions~\cite{Schechter:1980gr}. 
As a result, in this general neutrino framework, $U^{n\times n}$ can be expressed as the product of the new physics (NP) piece, times the Standard Model (SM) piece 
\begin{equation}
U^{n \times n}=U^{ NP } \, U^{SM} . 
\end{equation}
Thinking in terms of the seesaw mechanism it is convenient to express the full $U$ matrix as four submatrices.
Here we label them as in~\cite{Hettmansperger:2011bt}~\footnote{The form of the matrices N, S, T and V within the full seesaw expansion was given in Ref.~\cite{Schechter:1981cv}.
  They correspond, respectively, to $U_a$, $U_b$, $U_c$ and $U_d$ of Eqs.~(2.8) and (3.5) of the above reference.}, i.e.
\begin{equation}
U^{n\times n}=\left(\begin{array}{cc} N & S\\
  T & V
\end{array}\right) .
\end{equation}
Notice that we have a block, $N$, relating the light neutrino sector with the three active neutrino flavors.
Here $V$ will be a $(n-3)\times(n-3)$ submatrix, while $S$ and $T$ will be, in general, rectangular matrices. \\[-.4cm]

Clearly, in this general case,  the full unitarity condition will take the form
\begin{eqnarray}
 NN^\dagger + SS^\dagger = I , \nonumber \\
 TT^\dagger + VV^\dagger = I . 
\end{eqnarray}
Therefore, the $3\times3$ matrix $N$ describing the mixing of light neutrinos will no longer be unitary.
One can show~\cite{Escrihuela:2015wra} that in the most general case $N$ can  be parametrized as
\begin{equation}
  \label{eq:N}
N=N^{NP}\, U^{3\times3}=\left(\begin{array}{ccc}\alpha_{11} & 0 & 0\\
\alpha_{21} & \alpha_{22} & 0\\
\alpha_{31} & \alpha_{32} & \alpha_{33}
\end{array}\right)\: U^{3\times3},
\end{equation}
where the diagonal $\alpha$'s are real and close to 1, while the off-diagonals are small but complex.
Indeed, for any number of additional neutrino states, we will have for the diagonal entries of this matrix that
\begin{equation}
\alpha_{jj} \: 
= \: \prod_{i=4,n} \: \cos\theta_{ji} \, , 
\end{equation}
with no sum over $j$. For small mixings, the non-diagonal entries are given as
\begin{equation}
\alpha_{ji} \: \simeq\:
-\sum_{k=4,n} \theta_{jk}\theta_{ik}e^{-i(\varphi_{jk}-\varphi_{ik})} \,\,\, ; \,\,\,
i<j  \,\,\, .
\end{equation}
From this last expression, one can see that 
\begin{equation}
|\alpha_{21}|^2 
    \leq \sum^N_{i=4} |\theta_{2i}\theta_{1i}e^{-i(\varphi_{2i}-\varphi_{1i})} |^2 
     ~=~ \sum^N_{i=4} \theta^2_{2i}\theta^2_{1i}  , 
\end{equation}
and similar equations for the other two non-diagonal terms. Using the triangle inequality, one can now derive the consistency relations~\footnote{For a general derivation of this expression without the assumption of small mixing angles see~\cite{Forero:2021azc}.} 
,
\begin{equation}
|\alpha_{ji}| \le \sqrt{(1-\alpha_{jj}^2)(1-\alpha_{ii}^2)} .  \\
\end{equation}

The muon neutrino appearance probability will be given as  
%
%
\begin{eqnarray}
& P_{\mu e} = \alpha_{11}^2 |\alpha_{21}|^2 -4
        \sum\limits^3_{j>i} Re\left[ 
        N^*_{\mu j}N_{ej}N_{\mu i}N^*_{ei} \right]
        \sin^2\left(\frac{\Delta m^2_{ji}L}{4E}\right) \nonumber\\
&+ 
        2 \sum\limits^3_{j>i} Im\left[
        N^*_{\mu j}N_{ej}N_{\mu i}N^*_{ei}\right] 
        \sin\left(\frac{\Delta m^2_{ji}L}{2E}\right)  \,,
\end{eqnarray}
%
where $N$ is given in terms of the $\alpha$'s as in Eq.~(\ref{eq:N}).

For the case of vacuum oscillations, the parameters characterizing unitarity violation in the $\mu$-e sector are $\alpha_{11}$, $\alpha_{22}$, and $\alpha_{21}$. 
In the presence of matter effects, the appearance probability could also involve the third neutrino type, since the charged and neutral current potential will modify the effective form of the matrix $N$.
The charged current potential for the nonunitary case will be given by  
\begin{equation}
V^{\alpha \beta}_{CC} =
 \sqrt{2} \,G_{F} N_{e} \left(NN^{\dagger}\right)_{\alpha e}
    \left(NN^{\dagger}\right)_{e \beta} \, ,
  \end{equation}
where $G_F$ is the Fermi constant and $N_e$ the number density of electrons in the medium. The matrix product will be written, in terms of the $\alpha$'s, as~\cite{Escrihuela:2016ube}: 
\begin{equation}
(NN^\dagger)_{\alpha e}(NN^\dagger)_{e \beta}= \alpha_{11}^2
\left(
\begin{array}{ccc}
 \alpha _{11}^2 & \alpha _{11} \alpha _{21}^* & \alpha _{11} \alpha _{31}^* \\
 \alpha _{11} \alpha _{21} & |\alpha _{21}|^2 & \alpha _{21} \alpha _{31}^* \\
 \alpha _{11} \alpha _{31} & \alpha _{21}^* \alpha _{31} & |\alpha _{31}|^2 \\
\end{array}
\right)\, .
\end{equation}
The corresponding potential for neutral currents will be 
\begin{align}
V^{\alpha \beta}_{NC} =&-\sqrt{2}G_F \frac{N_n}{2}
\sum_{\rho}(NN^\dagger)_{\alpha \rho}(NN^\dagger)_{\rho
  \beta} \nonumber \\ 
  =&-\sqrt{2}G_F \frac{N_n}{2} \left[(NN^\dagger)^2
\right]_{\alpha \beta} \, ,
\end{align}
where the matrix product $(NN^\dagger)^2$, at leading order in the non-diagonal $\alpha$'s, takes the form~\cite{Escrihuela:2016ube} 
\begin{equation}
\resizebox{.49\textwidth}{!}{$
\left(
\begin{array}{ccc}
 \alpha_{11}^4 & \alpha_{11} \alpha^*_{21} \left(\alpha_{11}^2+\alpha_{22}^2\right) & \alpha_{11} \alpha^*_{31} \left(\alpha_{11}^2+\alpha_{33}^2\right) \\
 \alpha_{11} \alpha_{21} \left(\alpha_{11}^2+\alpha_{22}^2\right) & \alpha_{22}^4 & \alpha_{22} \alpha^*_{32} \left(\alpha_{22}^2+\alpha_{33}^2\right) \\
 \alpha_{11} \alpha_{31} \left(\alpha_{11}^2+\alpha_{33}^2\right) & \alpha_{22} \alpha_{32} \left(\alpha_{22}^2+\alpha _{33}^2\right) & \alpha_{33}^4 \\
\end{array}
\right) \,$}.
\end{equation}

Neglecting cubic terms in $\alpha_{21}$, $\sin\theta_{13}$, and $\Delta m^2_{21}$, one finds that, in the vacuum case limit, the main contribution to the conversion probability  will be given by 
\begin{equation}
\label{pme_vac}
P_{\mu e} = 
 (\alpha_{11}\alpha_{22})^2 P^{3\times3}_{\mu e}
+  \alpha_{11}^2 \alpha_{22}|\alpha_{21}|  P^{I}_{\mu e} 
+ \alpha_{11}^2|\alpha_{21} |^2 , 
\end{equation}
where $P^{3\times 3}_{\mu e}$ denotes the usual three-neutrino conversion probability,
%
\begin{align}
\tiny
P^{3\times 3}_{\mu e} &= 
 4 \bigg[\cos^2\theta_{12} \cos^2\theta_{23} 
   \sin^2\theta_{12} \sin^2\left(\frac{\Delta m^2_{21}L}{4E_\nu}\right) \nonumber \\
& +  \cos^2\theta_{13}\sin^2\theta_{13}
   \sin^2\theta_{23}\sin^2\left(\frac{\Delta m^2_{31}L}{4E_\nu}\right) \bigg] \nonumber\\ 
&+ 
  \sin 2\theta_{12}
   \sin\theta_{13}\sin 2\theta_{23}
   \sin\left(\frac{\Delta m^2_{21}L}{2E_\nu}\right)\times \nonumber \\
&  \sin\left(\frac{\Delta m^2_{31}L}{4E_\nu}\right) 
   \cos\left(\frac{\Delta m^2_{31}L}{4E_\nu} + \delta_\mathrm{CP}\right)  \,,  
\end{align}
%
where $P^{I}_{\mu e}$ is the interference term 
%
%
\begin{align}
P^{I}_{\mu e}  &= 
-2 
   \bigg[
   \sin(2\theta_{13}) \sin\theta_{23} 
   \sin\left( \frac{\Delta m^2_{31}L} {4E_\nu}\right)\times\nonumber \\
  & \hspace*{2.5cm}\sin\left(\frac{\Delta m^2_{31}L}{4E_\nu} + \delta_\mathrm{CP} - \phi_{21}\right) \bigg]
\nonumber \\ 
  & +   \cos\theta_{13} \cos\theta_{23} 
  \sin 2\theta_{12} \sin \phi_{21}
   \sin\left(\frac{\Delta m^2_{21}L}{2E_\nu}\right)
   ,
\end{align}
%
with $\phi_{21}=\text{arg}(\alpha_{21})$. 


\section{Three neutrino oscillation probabilities}
\label{sec:probability}
\begin{figure*}[t!]
\centering
\includegraphics[height=6.9cm,width=6.9cm]{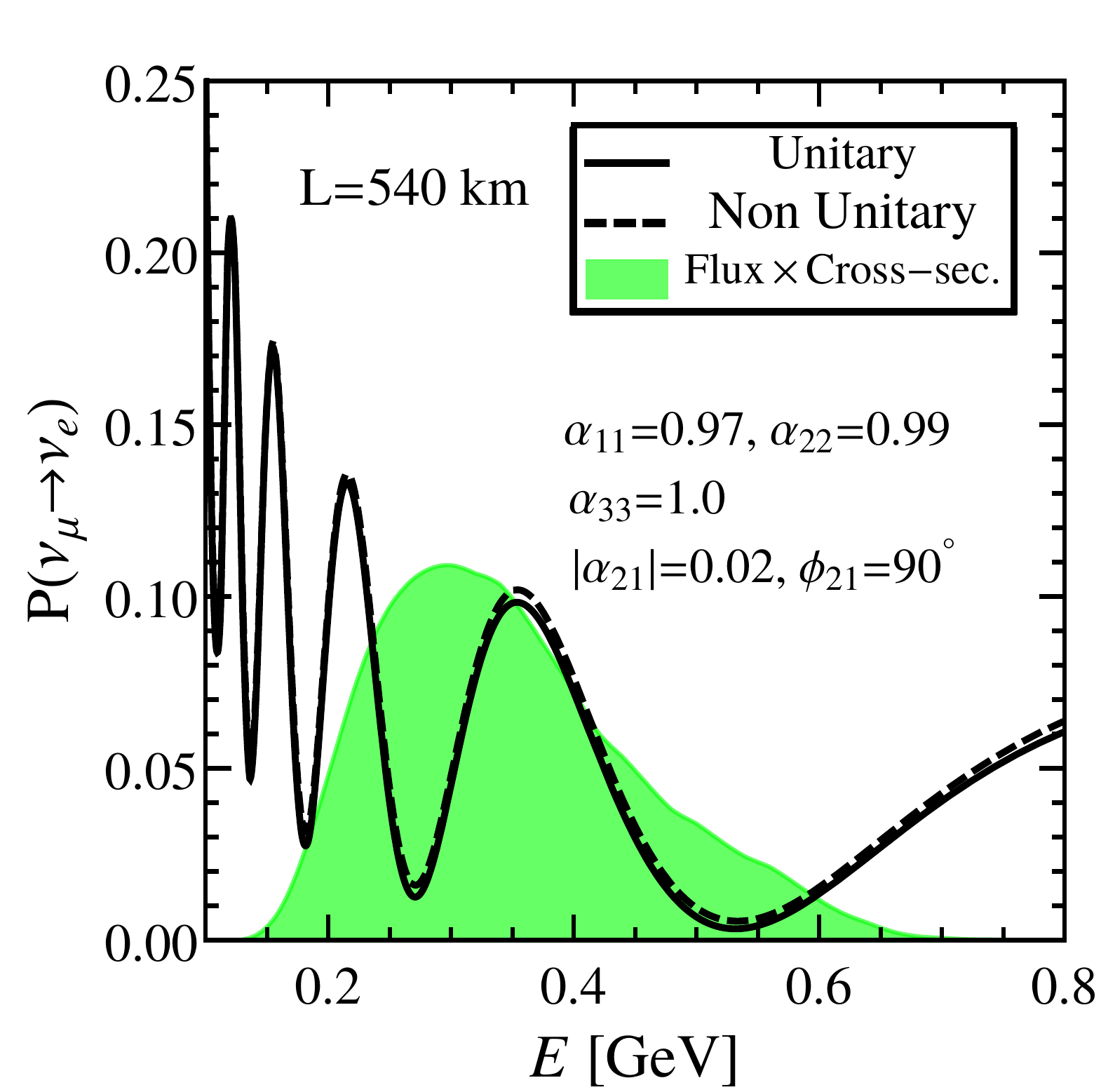}
\includegraphics[height=6.9cm,width=6.9cm]{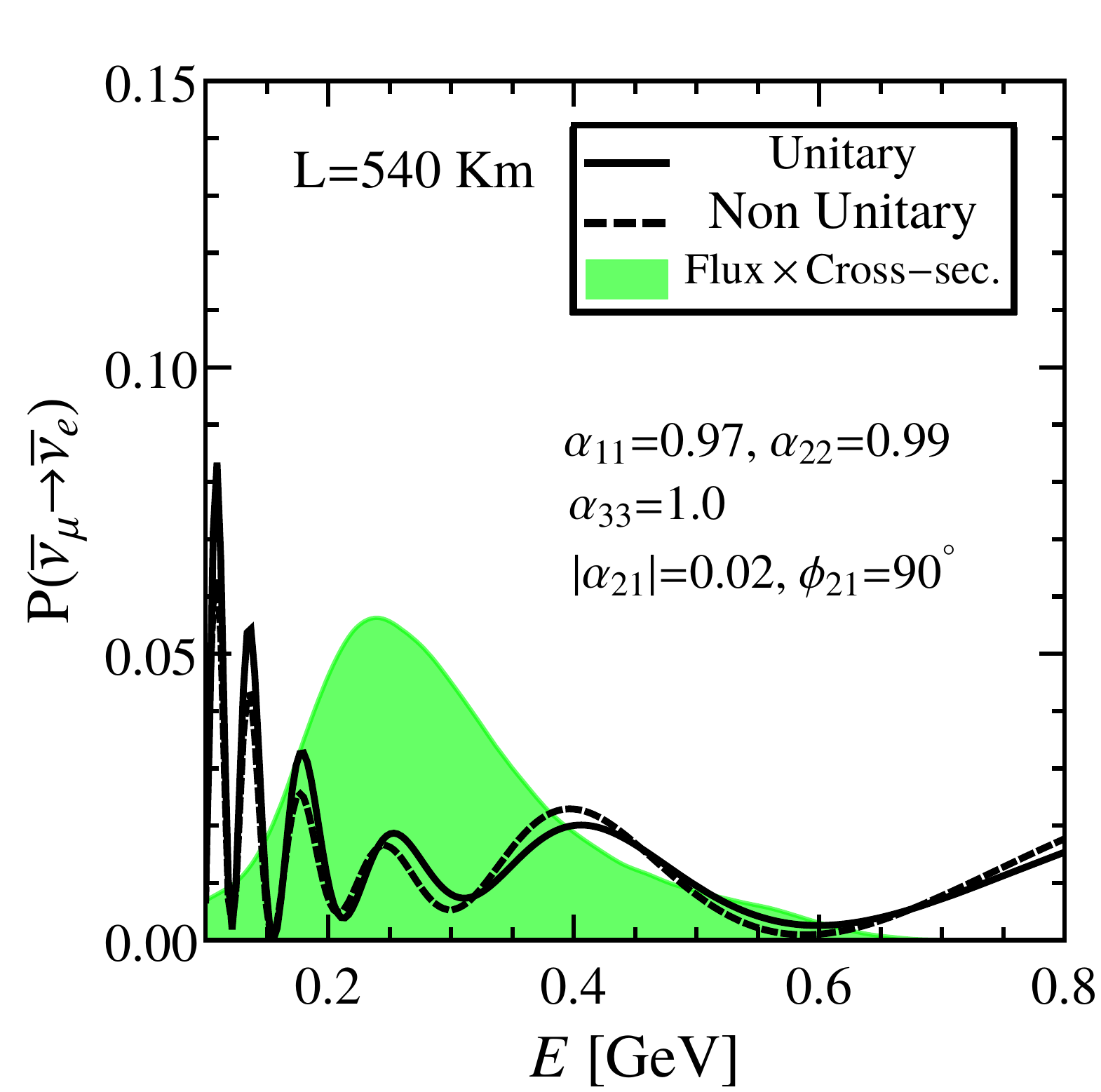}\\
\includegraphics[height=6.9cm,width=6.9cm]{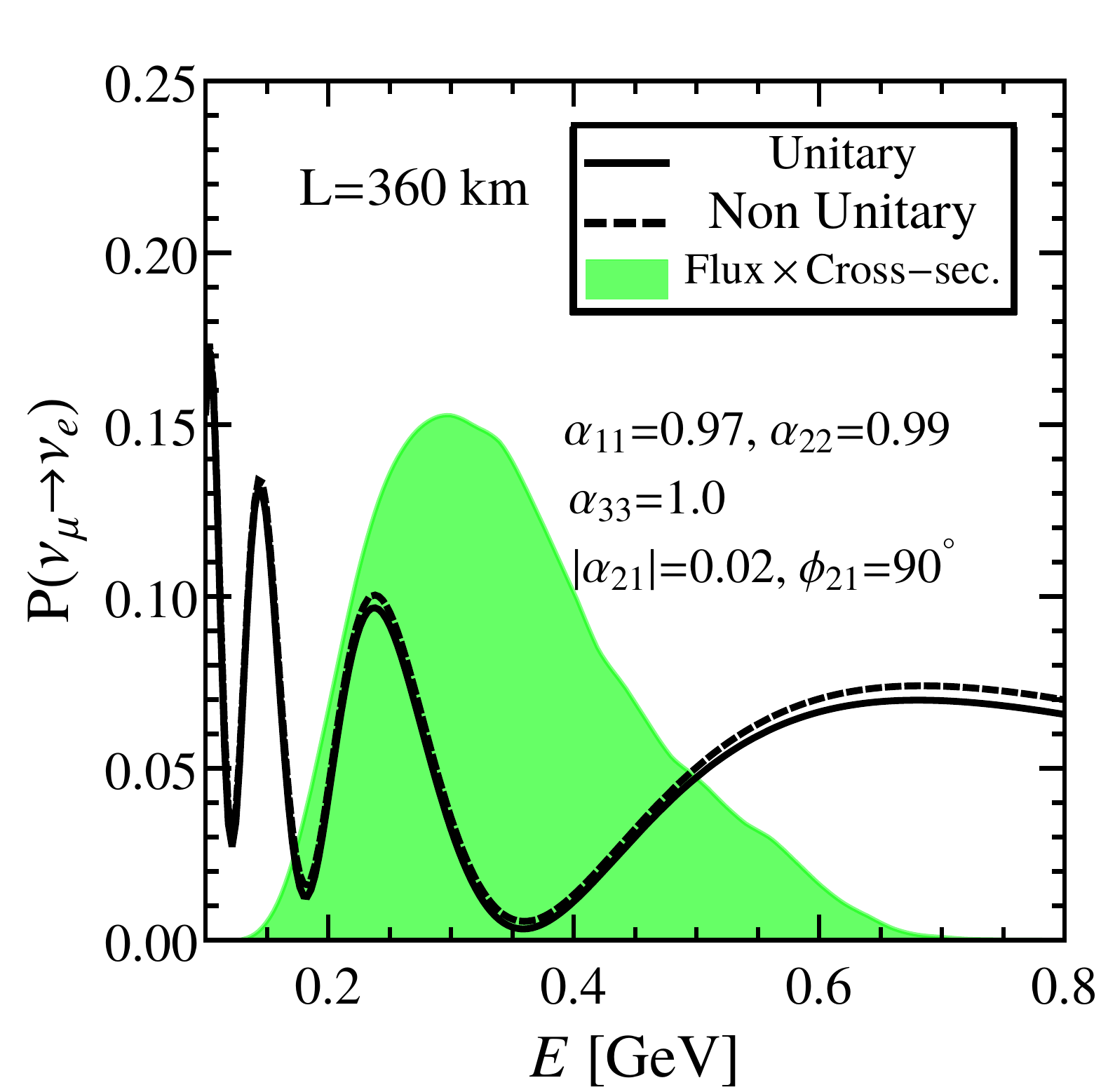}
\includegraphics[height=6.9cm,width=6.9cm]{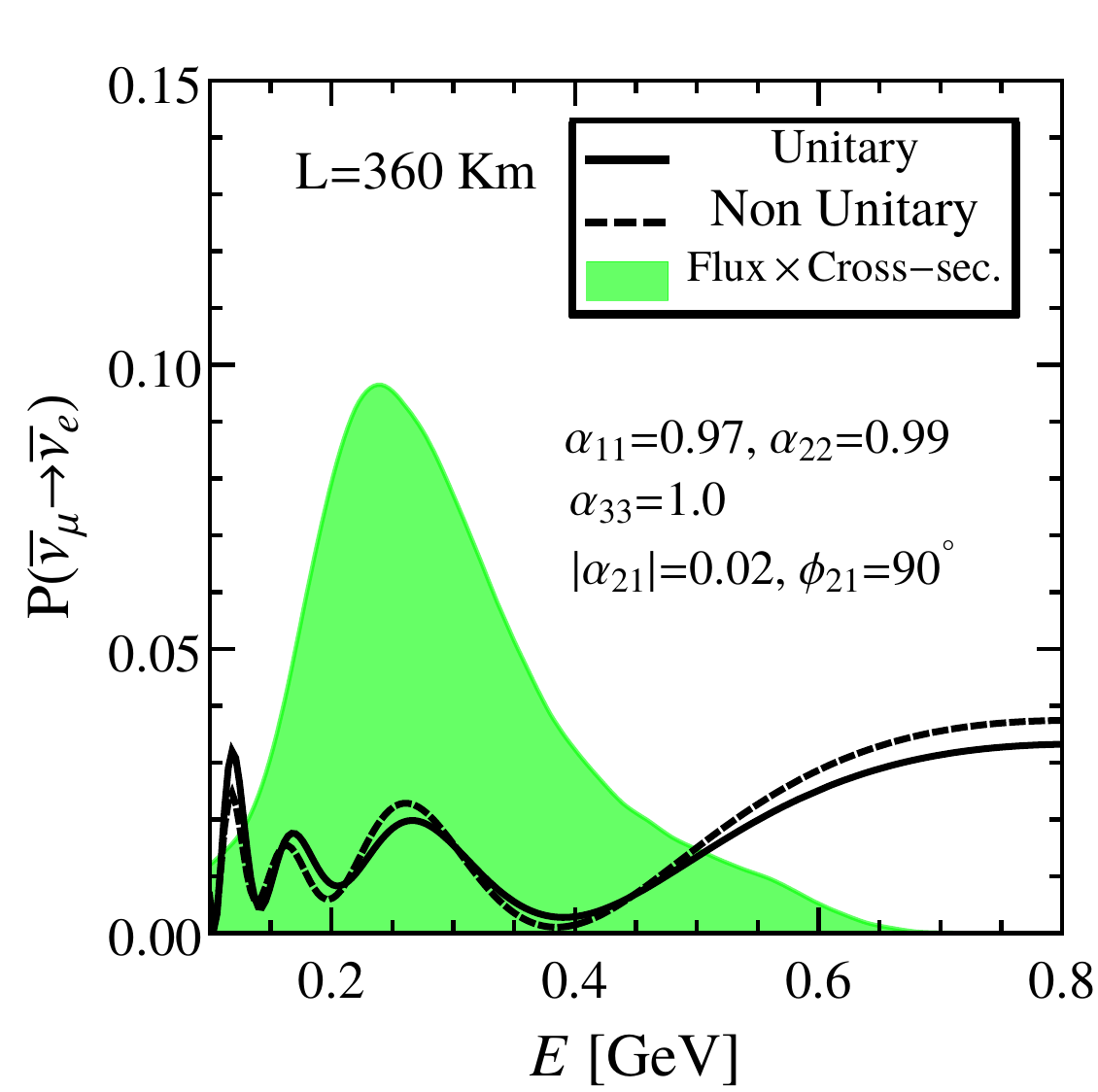}\\
\includegraphics[height=6.9cm,width=6.9cm]{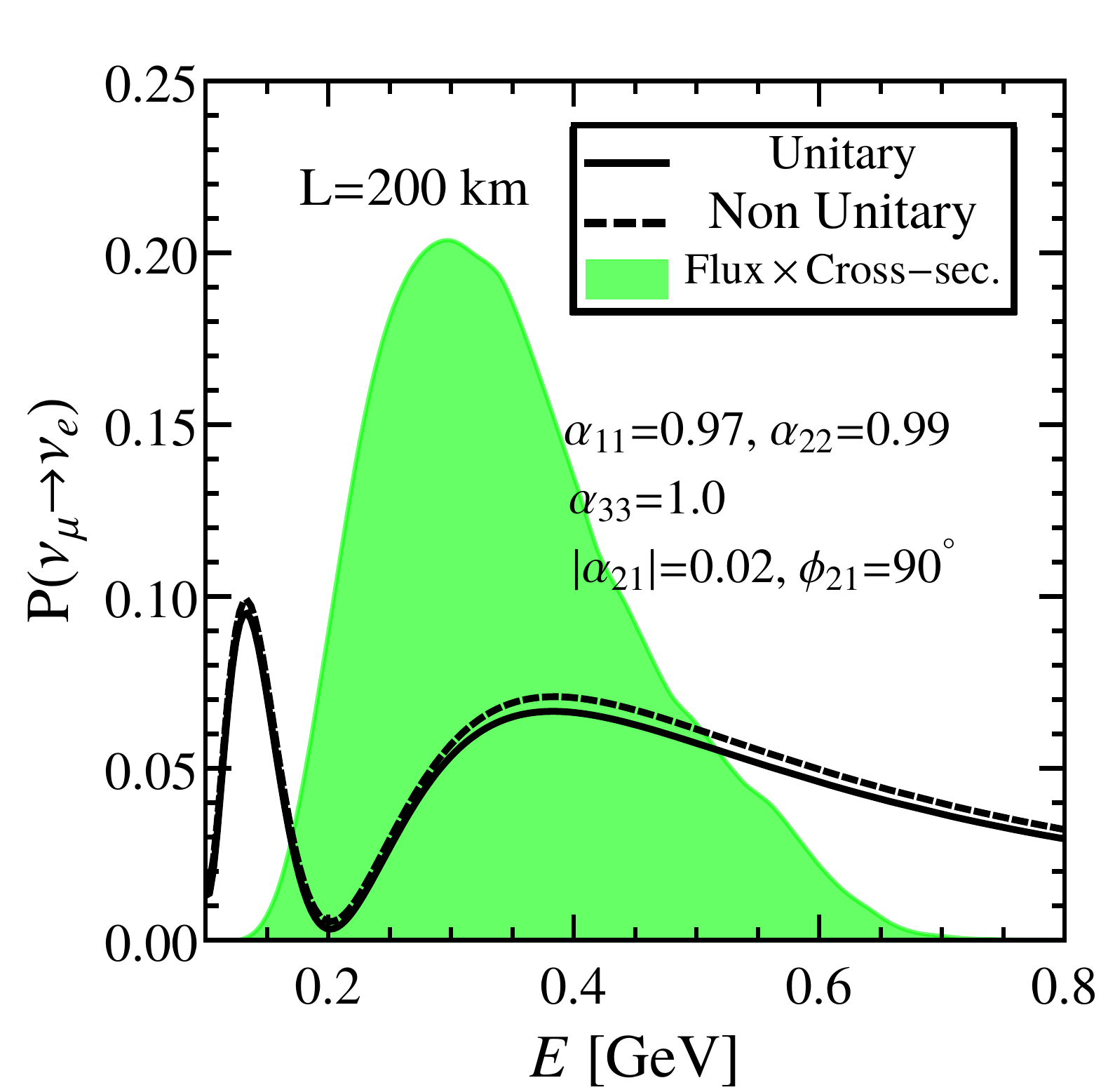}
\includegraphics[height=6.9cm,width=6.9cm]{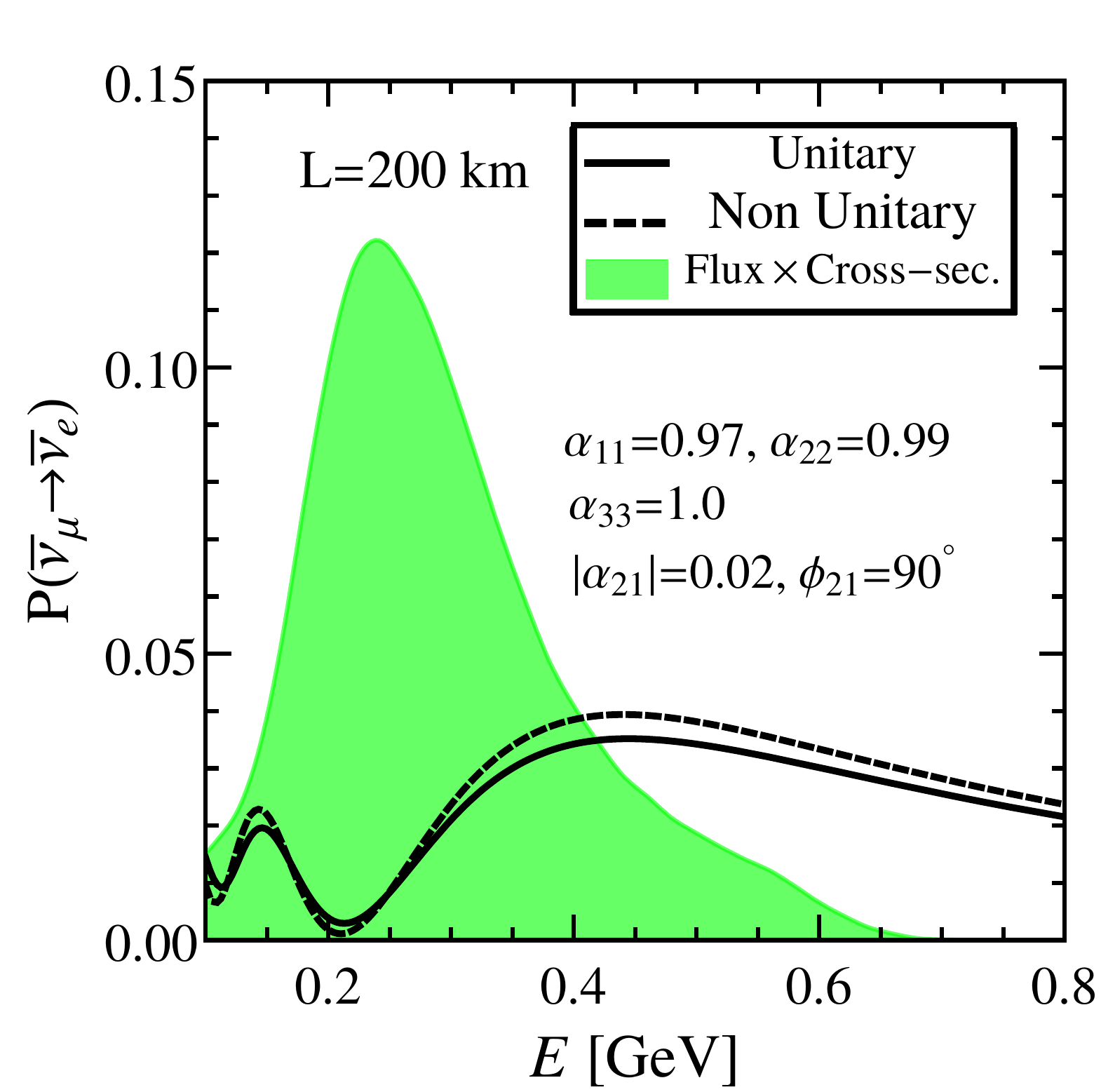}
\caption{The left panels represent the neutrino appearance probabilities for the unitary (solid line) and nonunitary (dashed line) oscillation case, for three different baselines, as indicated. The right panels show the corresponding antineutrino probabilities. The product of the $\nu_{\mu}$($\overline{\nu}_\mu$) fluxes and the $\nu_e$($\overline{\nu}_e$)-nucleus cross sections (in arbitrary units) is also shown as a green shaded area.} 
\label{fig:prob1}
\end{figure*}

Before coming to our numerical results, in this section we discuss the behaviour of the appearance (anti)neutrino probabilities. 
To this end, we show in the left (right) panel of Fig.~\ref{fig:prob1} the $\nu_{\mu} \to \nu_e$ ($\bar{\nu}_{\mu}\to  \bar{\nu}_e$) oscillation probabilities as a function of the neutrino energy. 
The upper, middle, and lower panels correspond to 540 km, 360 km, and 200 km baselines, respectively. 
We show the conversion probability in the standard unitary framework as a solid line, while the dashed one represents the nonunitary case. 
For the unitary case, we consider the values of the neutrino oscillation parameters given in Table~\ref{table1} with $\delta_{\rm CP} = -90^\circ$. 
For the nonunitary case, besides these values, we fix the nonunitary $\alpha$-parameters to be  $\alpha_{11}=0.97$, $\alpha_{22}=0.99$, $\alpha_{33}=1$, $|\alpha_{21}|=0.02$,
$\phi_{21}=90^\circ$, $|\alpha_{31}|=0$, and $|\alpha_{32}|=0$.
In Fig.~\ref{fig:prob1} we also include in green color (arbitrary units) the unoscillated (anti)neutrino $\nu_\mu$ flux times the (anti)$\nu_e$-nucleus cross section~\cite{Messier:1999kj,Paschos:2001np}.
  This provides a bird-eye view of the real analysis at the event level. The main region of interest extends from $0.1\, \rm GeV$ to $0.7 \,\rm GeV$, with a peak around $0.30\, \rm GeV$.
One can see the relative location of this peak with respect to the second oscillation maximum for the three baselines of $540$~km and $360$~km and the $200$~km case, see also Table~\ref{table2}. 
The true expected event number for the appearance signal must, of course, take into account the convolution of the cross section,
  the appearance probability and the unoscillated neutrino energy spectrum at the detector, which depends on its distance to the source. Thus, the total unoscillated neutrino flux for the $200$~km case will be approximately three times that for a $360$~km baseline, and seven times the flux of the $540$~km case. As a result, the shortest the baseline, the higher the expected number of events, as we will see in Fig.~\ref{events_spectra}.  

\begin{table*}[t!]
{%
\begin{center}
\begin{adjustbox}{width=0.95\textwidth}
\begin{tabular}{|c|c|c|c|c|c|c|}
\hline
$3\nu$ parameters & $\sin^2\theta_{12}$ & $\sin^2\theta_{13}$ & $\sin^2\theta_{23}$ & $\delta_{\rm CP}$ & $\frac{\Delta m_{21}^2}{10^{-5}\,\rm{eV^2}}$ & $\frac{\Delta m_{31}^2}{10^{-3}\,\rm{eV^2}}$ \\ 
\hline
Benchmark values & $0.318$ & $0.022$ & $0.57$ & $[-180,\,180]$ & $7.5$ & $2.55$ \\ 
\hline
Status   & fixed & marginalized & marginalized & marginalized & fixed & marginalized \\ 
 \hline
\end{tabular}
\end{adjustbox}
\end{center}
}%
\caption{Benchmark values for the standard three-neutrino oscillation parameters taken from the current global fit analysis~\cite{deSalas:2020pgw},
along with their assumed marginalization status within our analyses, except $\delta_{\rm CP}$. For simplicity, normal mass ordering (NO) has been assumed. }
\label{table1}
\end{table*}

\begin{table}[h!]
{%
\begin{center}
\begin{tabular}{|c|c|c|}
\hline
 baseline (km) & 1st osc. max. (GeV) & 2nd osc. max. (GeV)  \\
\hline
540 &1.05 & 0.35  \\
\hline
360 & 0.70 & 0.23 \\
\hline
200 & 0.39 & 0.13 \\
\hline
\end{tabular}
\end{center}
}%
\caption{Possible baselines of the ESSnuSB project~\cite{Baussan:2013zcy,Dracos:2016wso,Dracos:2018jsn,Dracos:2018syh}, along with the corresponding values of the neutrino energy for the first and second oscillation maxima. }
\label{table2}
\end{table}

\section{Experimental setup and simulation}
\label{sec:experimental-setup}

In this section, we briefly discuss the experimental specifications of the ESSnuSB and DUNE setups used in this work, followed by a description of our simulation procedure. 

\subsection{Experimental setup options}
\label{sec:experimental-setup-1}

The ESSnuSB project is a proposed accelerator neutrino experiment sourced at Lund (Sweden), where the ESS linac facility is currently under construction. 
The original ESSnuSB proposal was to use of a very intense proton beam of $2$~GeV energy and an average beam power of 5 MW, resulting in $2.7\times 10^{23}$ protons on target (POT) per year ($208$ effective days)~\cite{Baussan:2013zcy,Dracos:2016wso,Dracos:2018jsn,Dracos:2018syh}.
Here we will adopt this configuration. It is expected that the future linac upgrade can increase the proton energy up to 3.6 GeV. The neutrino and antineutrino fluxes arising from the 2 GeV proton beam peak around $0.25$ GeV~\cite{enrique}.
  These (anti)neutrinos will be detected by a 500 kton fiducial mass Water Cherenkov detector similar to the MEMPHYS project~\cite{Agostino:2012fd,luca}.
  Since the baseline of the far detector has not been finalized yet, we have considered in this work three possible baselines~\cite{Baussan:2013zcy}, which are 200 km, 360 km, and 540 km respectively. 
  It has been shown in~\cite{Baussan:2013zcy} that if the detector is placed in any of the existing mines in between 200 km to 600 km from the ESSnuSB site Lund, a $3\sigma$ evidence of CP violation could be achieved for 60\% coverage of the full $\delta_{\rm CP}$ range. 
Our simulation matches the event numbers of Table 3 and all other results given in~\cite{Baussan:2013zcy}.
  In all the numerical results presented here, we have assumed $2$~years of neutrino and $8$~years of antineutrino running with an optimistic assumption of uncorrelated $5$\% signal normalization and 10\% background normalization error for both neutrino and antineutrino appearance and disappearance channels, respectively. 
  For more details about the accelerator facility, beamline design, detector and baseline positions of this setup, see~\cite{Baussan:2013zcy}.
  Note that, while working on this paper, an updated analysis from the collaboration has come out~\cite{Alekou:2021coj}~\footnote{Ref.~\cite{Alekou:2021coj} has not considered
    the 200 km baseline option which, as we will see, actually turns out to be somewhat more promising for certain tasks, e.g. probing $\alpha_{21}$.}
Enhanced sensitivities to unitarity violation might be expected for the updated setup of the proposal. 
 This highly potential and ambitious facility is expected to start taking data around the year 2030. \\[-.3cm]
 
 DUNE is a future long-baseline accelerator-based neutrino experiment with a baseline of 1300 km from the source at Fermilab to the far detector placed deep underground at the Sanford Laboratory site in South Dakota.
DUNE will use a 40 kton LArTPC detector and a 120 GeV proton beam with 1.2 MW beam power resulting in $1.1\times 10^{21}$ POT/year.
For the numerical simulations, we have followed the experimental configurations provided by the collaboration in the Technical Design Report (TDR)~\cite{Abi:2020qib,Abi:2021arg},
assuming equal runtime of 3.5 years in neutrino and antineutrino mode, which results in 336~kton-MW-year exposure for the TDR setup.
 More details on the systematic errors, efficiencies and energy resolutions can be found in Refs.~\cite{Alion:2016uaj, Abi:2021arg}.

\subsection{Simulation procedure}
\label{sec:simulation-procedure}

In order to assess the statistical sensitivity of the ESSnuSB facility to neutrino oscillations, we have made use of the built-in $\chi^2$ function of the GLoBES package~\cite{Huber:2004ka,Huber:2007ji},
which incorporates the systematic errors through the pull terms~\cite{Huber:2002mx}. To perform the nonunitarity analysis we have used the modified  version of~\cite{Kopp_NSI}.
The total $\chi^2$ is a sum of all the contributions coming from different channels, 
\begin{eqnarray}
\underset{total}{\chi^2} = \underset{\nu_{\mu} \to \nu_e}{\chi^2} + \underset{\bar{\nu}_{\mu} \to \bar{\nu}_e}{\chi^2} +\underset{\nu_{\mu} \to \nu_{\mu}}{\chi^2} + \underset{\bar{\nu}_{\mu} \to \bar{\nu}_{\mu}}{\chi^2}.
\end{eqnarray}
Unless stated otherwise, the benchmark choices of the standard three-neutrino unitary oscillation parameters and their marginalization status in our analysis are given in Table~\ref{table1}. 
Our benchmark choices closely follow the current global fit analysis~\cite{deSalas:2020pgw}. 
Following the same analysis, we have adopted a $1\%$ uncertainty on the atmospheric mass-squared splitting $\Delta m_{31}^2$ and a $3.2\%$ uncertainty on the reactor mixing angle $\sin^2\theta_{13}$. 
We have freely marginalized over the atmospheric parameter $\sin^2\theta_{23}$  from $0.35$ to $0.65$.\\[-.3cm]

For the case of ESSnuSB, we have considered a line-averaged constant matter density $\rho = 2.8\,\rm g/cm^3$ following the PREM profile~\cite{DZIEWONSKI1981297, stacey:1977}. 
For DUNE, we have also assumed the same matter density but with a $5\%$ uncertainty due to the longer baseline.
For definiteness, we have assumed the currently preferred case of normal mass ordering (NO) throughout all of our analyses. 
Whenever appropriate, we have also marginalized over the NU parameters, along with their associated CP phases, implementing the current $3\sigma$ bounds shown in Table~\ref{table3}.
These come essentially from short-baseline oscillation searches such as from NOMAD~\cite{NOMAD:2001xxt,NOMAD:2003mqg} and CHORUS~\cite{CHORUS:1997wxi,CHORUS:2007wlo}
  and the long-baseline experiments T2K~\cite{T2K:2021xwb}, NOvA~\cite{NOVA2020} and, most importantly, MINOS/MINOS+~\cite{MINOS:2017cae}.

\begin{table}[h!]
{%
\begin{center}
\begin{adjustbox}{width=.48\textwidth}
\begin{tabular}{|c|c|c|c|c|c|c|}
\hline
 NU parameters & $|\alpha_{21}|$ & $|\alpha_{31}|$ & $|\alpha_{32}|$  & $\alpha_{11}$ & $\alpha_{22}$ & $\alpha_{33}$ \\ 
\hline
3$\sigma$ bounds   & $< 0.025$ & $< 0.077$ & $<  0.020$  & $>  0.929$ & $>  0.987$ & $>  0.715$ \\
\hline
\end{tabular}
\end{adjustbox}
\end{center}
}%
\caption{ 
Current neutrino constraints on the nonunitary parameters from Ref.~\cite{Forero:2021azc}.} 
\label{table3}
\end{table}

\section{Results}
\label{sec:results}

In this section we discuss in detail the numerical findings of our analyses, where we explore the sensitivity of the ESSnuSB facility to the nonunitary neutrino mixing,
as well as the impact of nonunitarity on the measurement of the standard three-neutrino oscillation parameters, with emphasis on the CP-violating phase, $\delta_{\rm CP}$.  

\subsection{Probing nonunitary neutrino mixing at ESSnuSB}
\label{sec:essn-sens-non}
%
\begin{figure*}[t]
\centering
\includegraphics[width=1.05\textwidth]{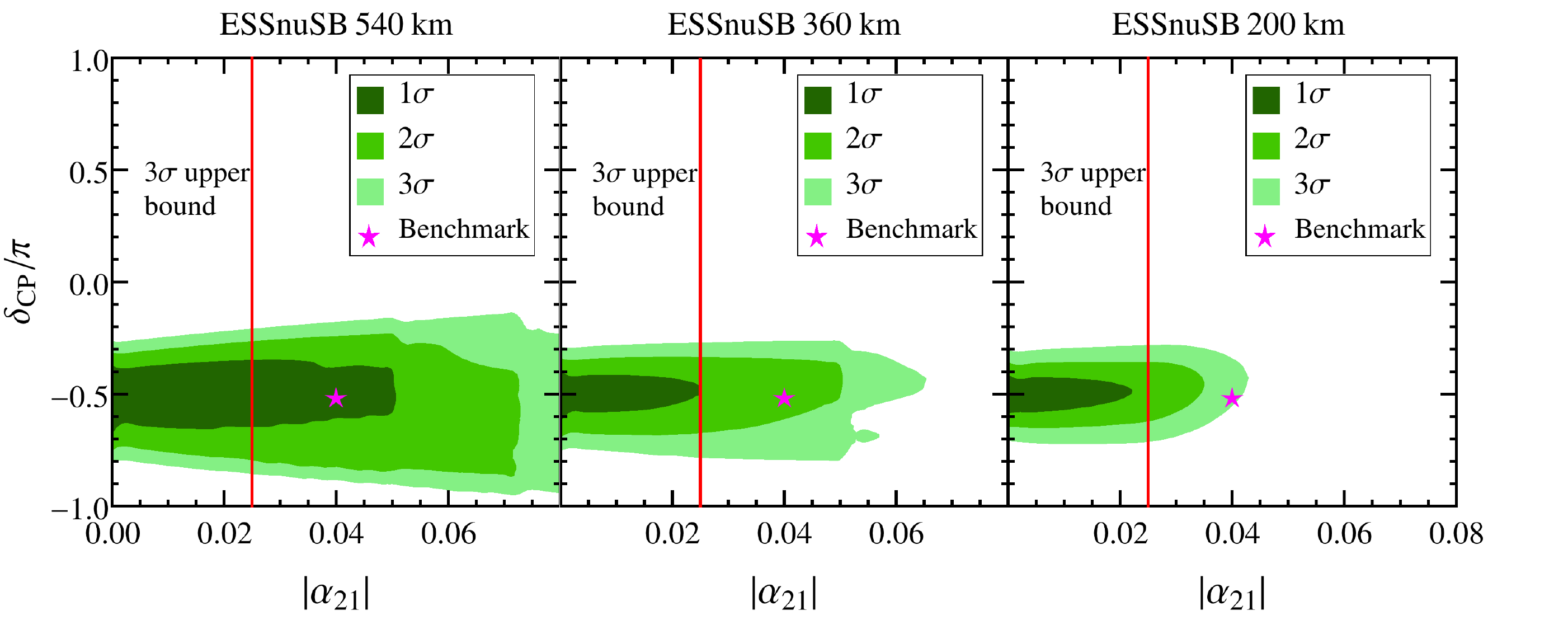}
\caption{{ESSnuSB sensitivity to the nonunitarity scenario} in the   $(|\alpha_{21}|,\, \delta_{CP})$ plane. Left, middle, and right panels correspond to 540, 360, and 200 km baselines, respectively. 
Contours are shown at $1\sigma$, $2\sigma$, and $3\sigma$  for 2 d.o.f.
The red vertical lines indicate the current $3\sigma$ upper limit on $|\alpha_{21}|$ from neutrino data. 
We have assumed $\delta_{\rm CP}(\rm true) = -90^\circ$ and normal mass ordering. The magenta star marked as benchmark in each panel has been used later to produce dashed spectra in the upper panel of Fig.~\ref{events_spectra}.} 
\label{alpha21_phi21}
\end{figure*} 
We start our discussion from Fig.~\ref{alpha21_phi21}, where we show the ESSnuSB sensitivity to nonunitarity in the 
 $(|\alpha_{21}|,\,  \delta_{\rm CP})$ plane. 
Left, middle and right panels show the results for 540, 360, and 200 km baselines, respectively. 
The light, medium, and dark green contours in each panel correspond to the $1\sigma$, $2\sigma$, and $3\sigma$ allowed regions for 2 degrees of freedom (d.o.f.) i.e., $\Delta\chi^2 = 2.3,\,6.18,$ and $11.83$, respectively. 
In this figure we have assumed the standard unitary framework as the true hypothesis, and then we have fitted the nonunitary hypothesis against it. 
Normal mass ordering has been assumed all over the analyses. 
The true data have been generated assuming the benchmark choices of the standard unitary oscillation parameters in Table~\ref{table1} with $\delta_{\rm CP}(\rm true) = -90^\circ$, while for the reconstruction we have fixed the solar oscillation parameters and marginalized over $\theta_{13}$, $\theta_{23}$, and $\Delta m_{31}^2$.
In addition, we have also marginalized over the NU parameters $\alpha_{11}$, $\alpha_{22}$, and $\alpha_{33}$ within their allowed $3\sigma$ ranges as given in Table~\ref{table3}. 
We have also freely varied $\phi_{21}$ from $-\pi$ to $+\pi$.
We have assumed zero values of the other non-diagonal NU parameters, $|\alpha_{31}|$ and $|\alpha_{32}|$. 
The red vertical lines in each panel indicate the current $3\sigma$ upper limit on $|\alpha_{21}|$ from  neutrino data. 
One sees that the 200 km baseline gives the best $1\sigma$, $2\sigma$, and $3\sigma$ sensitivities on $|\alpha_{21}|$, in comparison to the $360$ km and $540$ km baseline option. 
 Quantitatively, the attainable upper limits for $|\alpha_{21}|$ at $1\sigma$ are $0.044,\, 0.024,\,\rm{and}\,\, 0.02$ for 540 km, 360 km, and 200 km baselines, respectively. 
Conversely, we have checked that, there is basically no sensitivity to the new CP phase $\phi_{21}$ for any of the baselines. 
On the other hand, the measurement on $\delta_{\rm CP}$ is not affected much by the presence of nonunitarity.  
One can also see that the uncertainty on the measurement of $\delta_{\rm CP}$ is the lowest for 200 km and the highest for 540 km. 
\begin{figure*}[t]
\hspace{.5cm}
\includegraphics[width=1\textwidth]{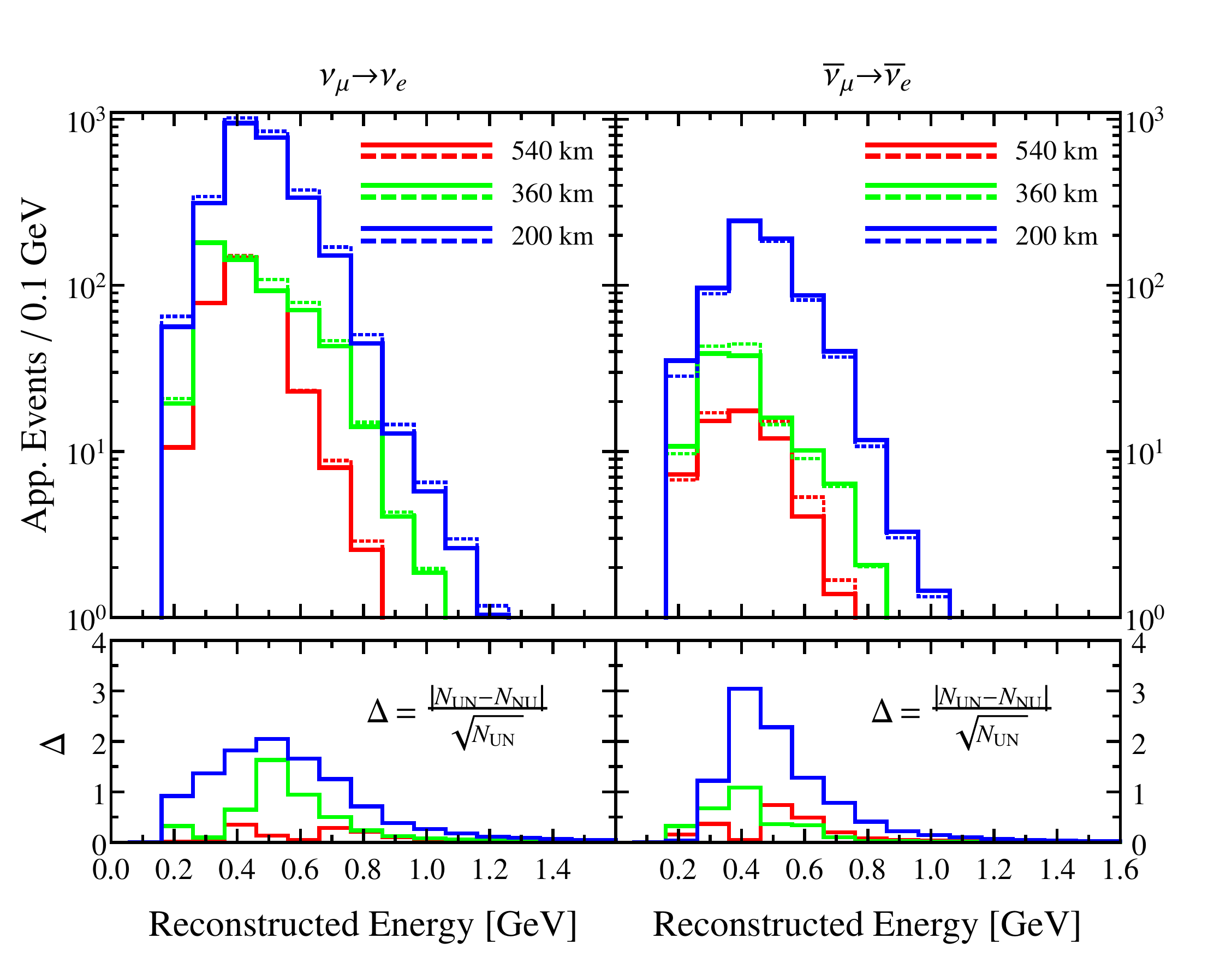}
\caption{
  Upper panels:  $\nu_{\mu}\to  \nu_e$ and $\bar{\nu}_{\mu}\to  \bar{\nu}_e$ appearance event-spectra as a function of reconstructed neutrino energy at ESSnuSB for different baselines, as indicated.
  Solid lines represent the standard three-neutrino spectra generated with the standard oscillation parameter values from Table~\ref{table1} with $\delta_{\rm  CP} = -90^\circ$ and dashed lines correspond to the NU scenario taking the optimized values of the standard oscillation parameters (not shown explicitly) and the new physics nonunitarity parameters from Table~\ref{table4}.
  The lower panels represent the absolute difference in the appearance events between the standard unitary  (UN) and the NU scenarios divided by the statistical uncertainty in each bin.}
\label{events_spectra}
\end{figure*}

%
%
\begin{table}[h!]
{%
\begin{center}
\begin{adjustbox}{width=.48\textwidth}
\begin{tabular}{|c|c|c|c|c|c|c|c|}
\hline
baseline (km)& \quad $\alpha_{11}$ \quad & \quad $\alpha_{22}$ \quad & \quad $\alpha_{33}$ \quad & \quad $\phi_{21}$ \quad & \quad $|\alpha_{31}|$ \quad &\quad  $|\alpha_{32}|$ \quad \\ 
\hline
$540$ &   $0.95$  &  $0.99$  &  $0.73$  & $155^\circ$  & $0$ & $0$ \\ 
\hline
$360$ &  $0.94$  &  $0.99$  &  $0.73$  & $140^\circ$ & $0$ & $0$ \\ 
\hline
$200$ &  $0.94$  &  $0.99$  &  $0.96$  & $75^\circ$ & $0$ & $0$ \\ 
 \hline
\end{tabular}
\end{adjustbox}
\end{center}
}%
\caption{Nonunitarity parameters used for obtaining the nonunitary spectra (dashed line) in Fig.~\ref{events_spectra}, see text for more details. }
\label{table4}
\end{table}
%

\begin{figure*}[t!]
\centering
\includegraphics[height=5.5cm,width=5.5cm]{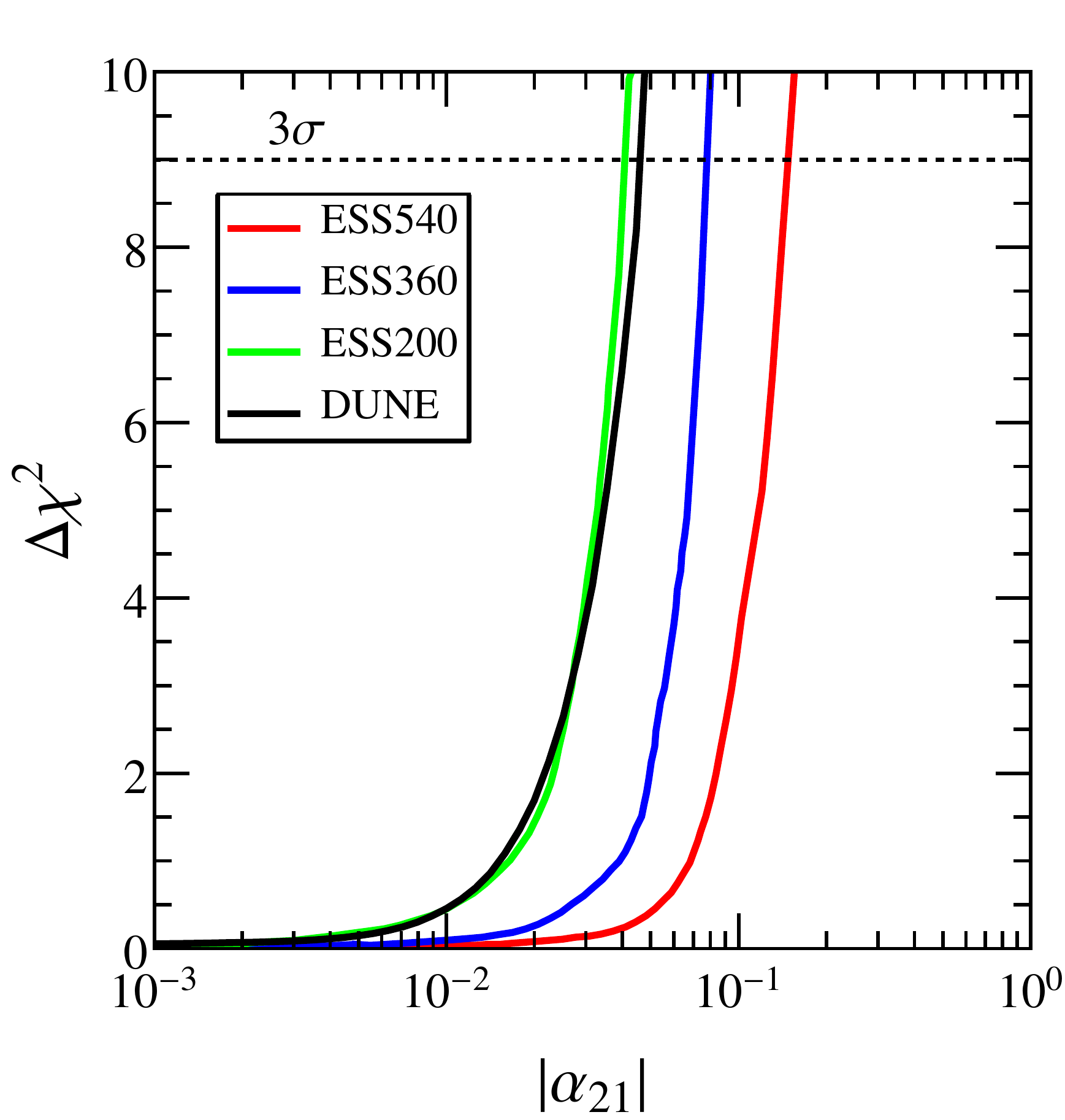}
\includegraphics[height=5.5cm,width=5.5cm]{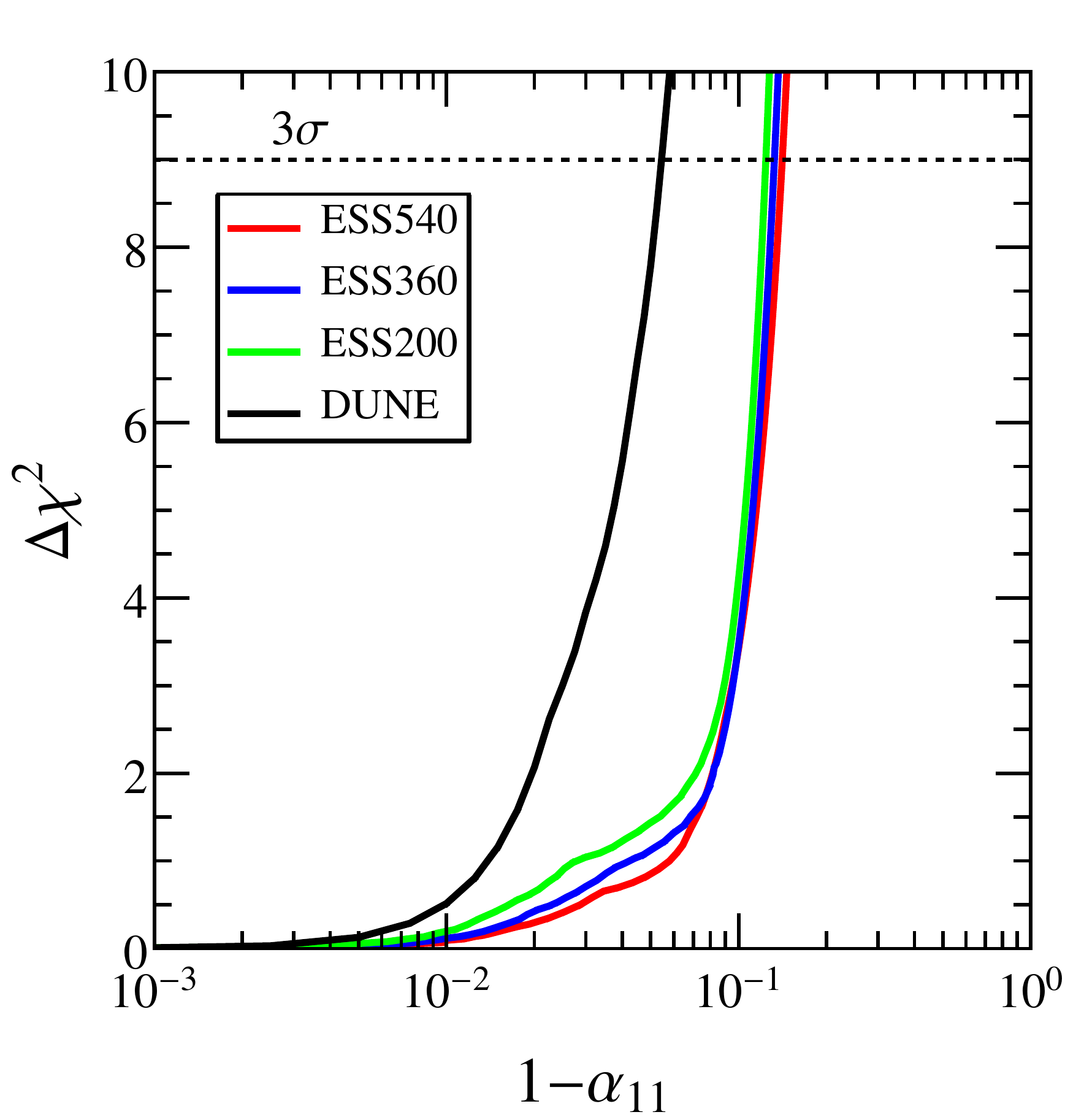}
\includegraphics[height=5.5cm,width=5.5cm]{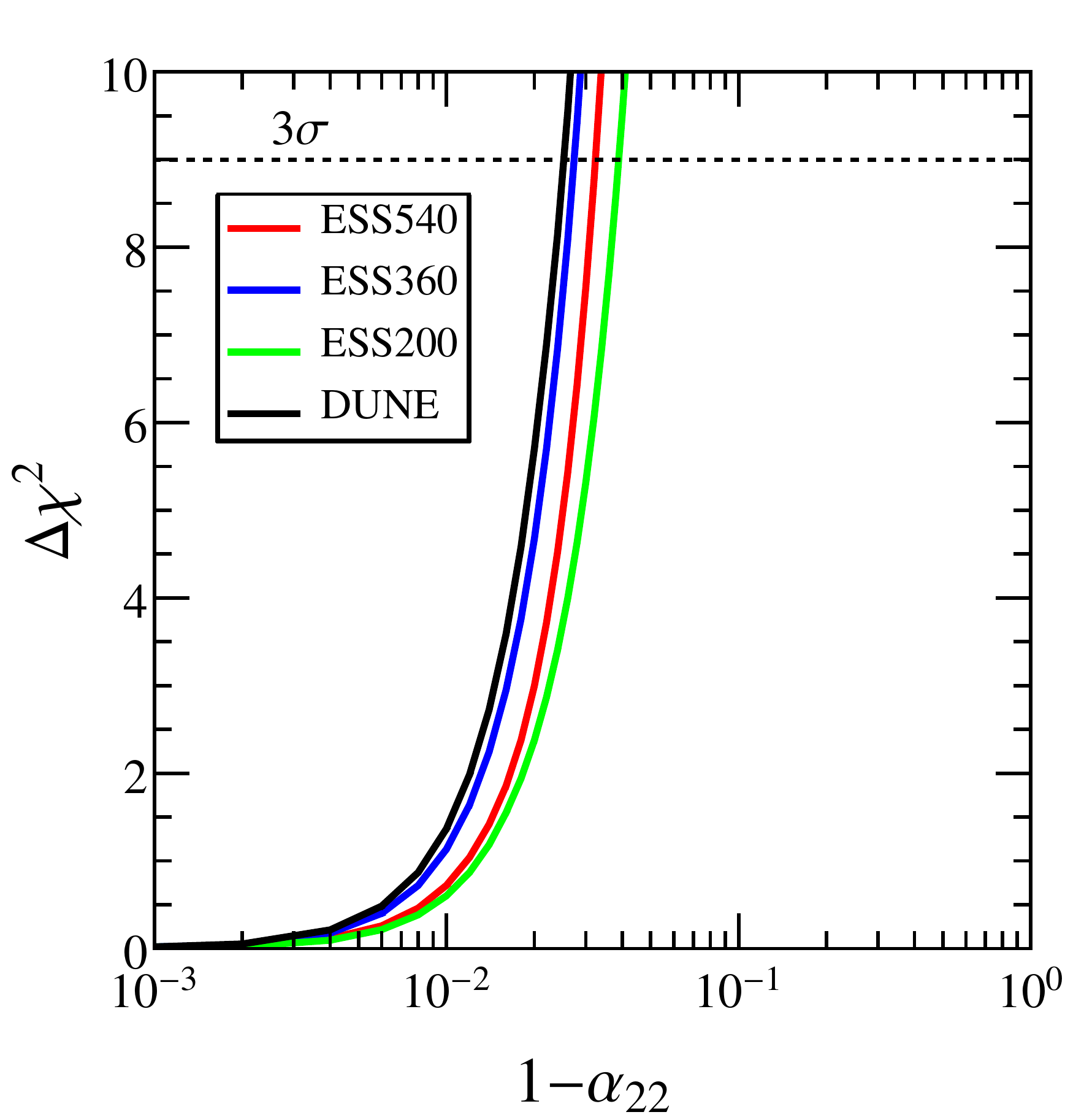}\\
\includegraphics[height=5.5cm,width=5.5cm]{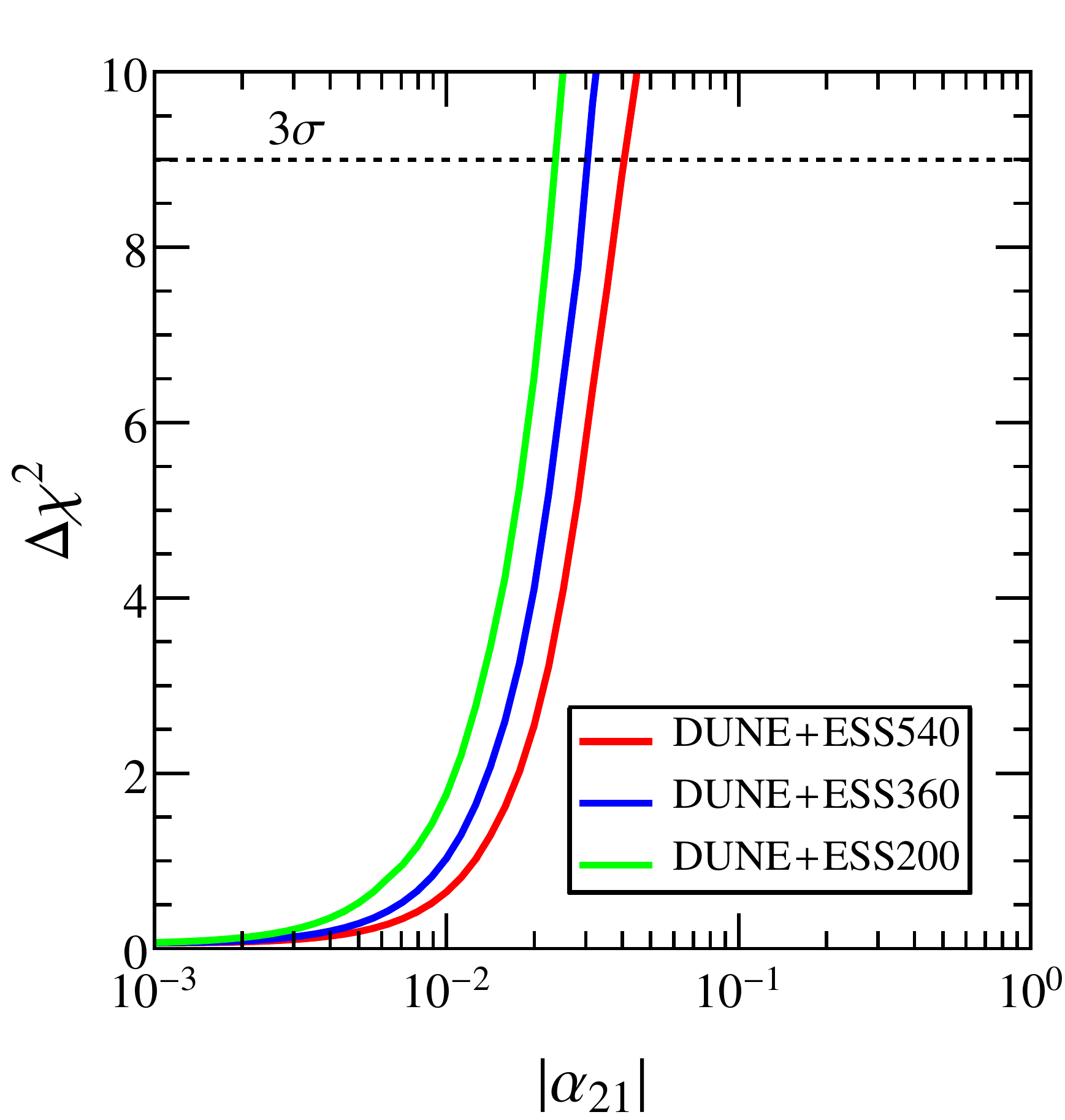}
\includegraphics[height=5.5cm,width=5.5cm]{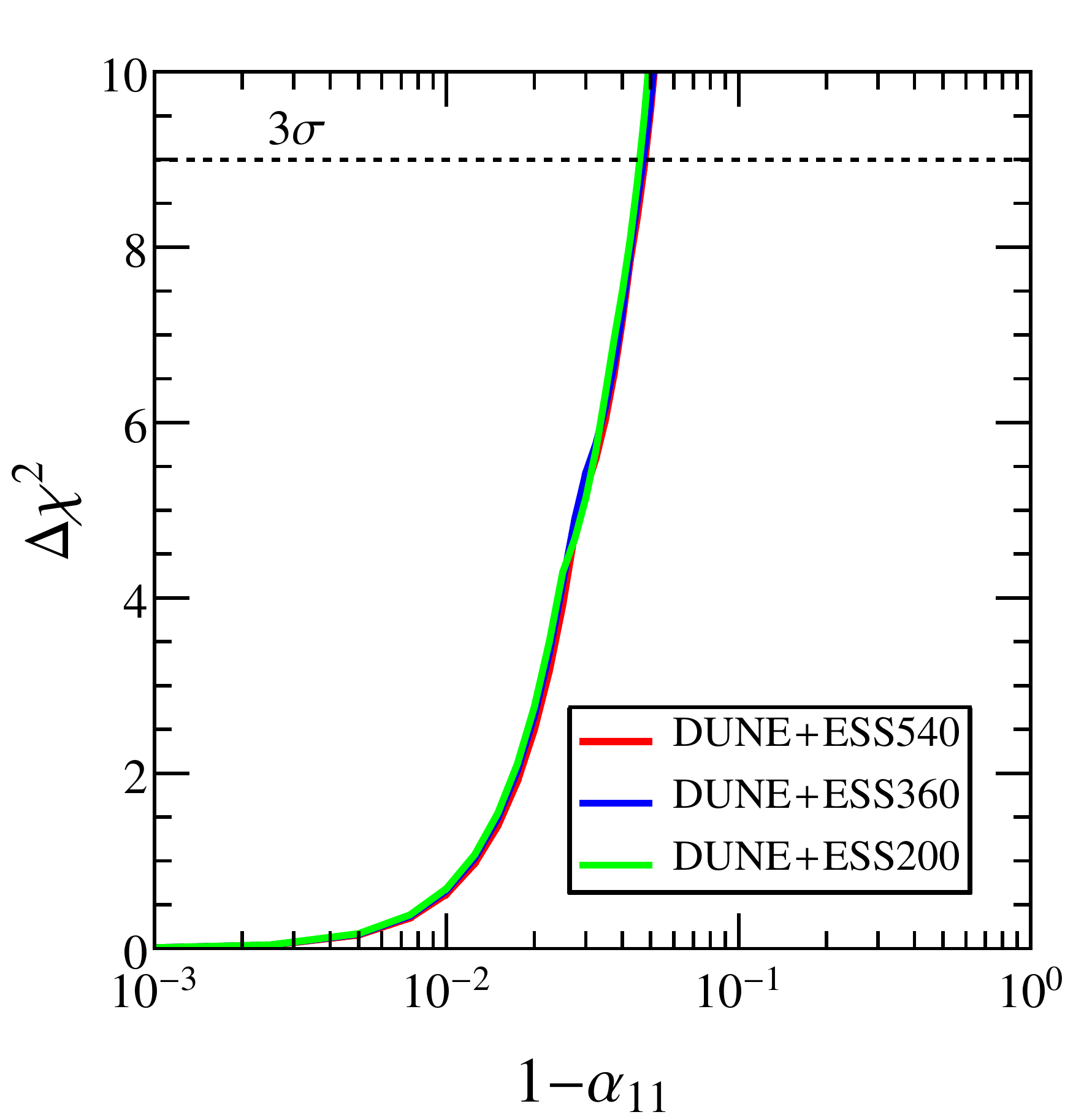}
\includegraphics[height=5.5cm,width=5.5cm]{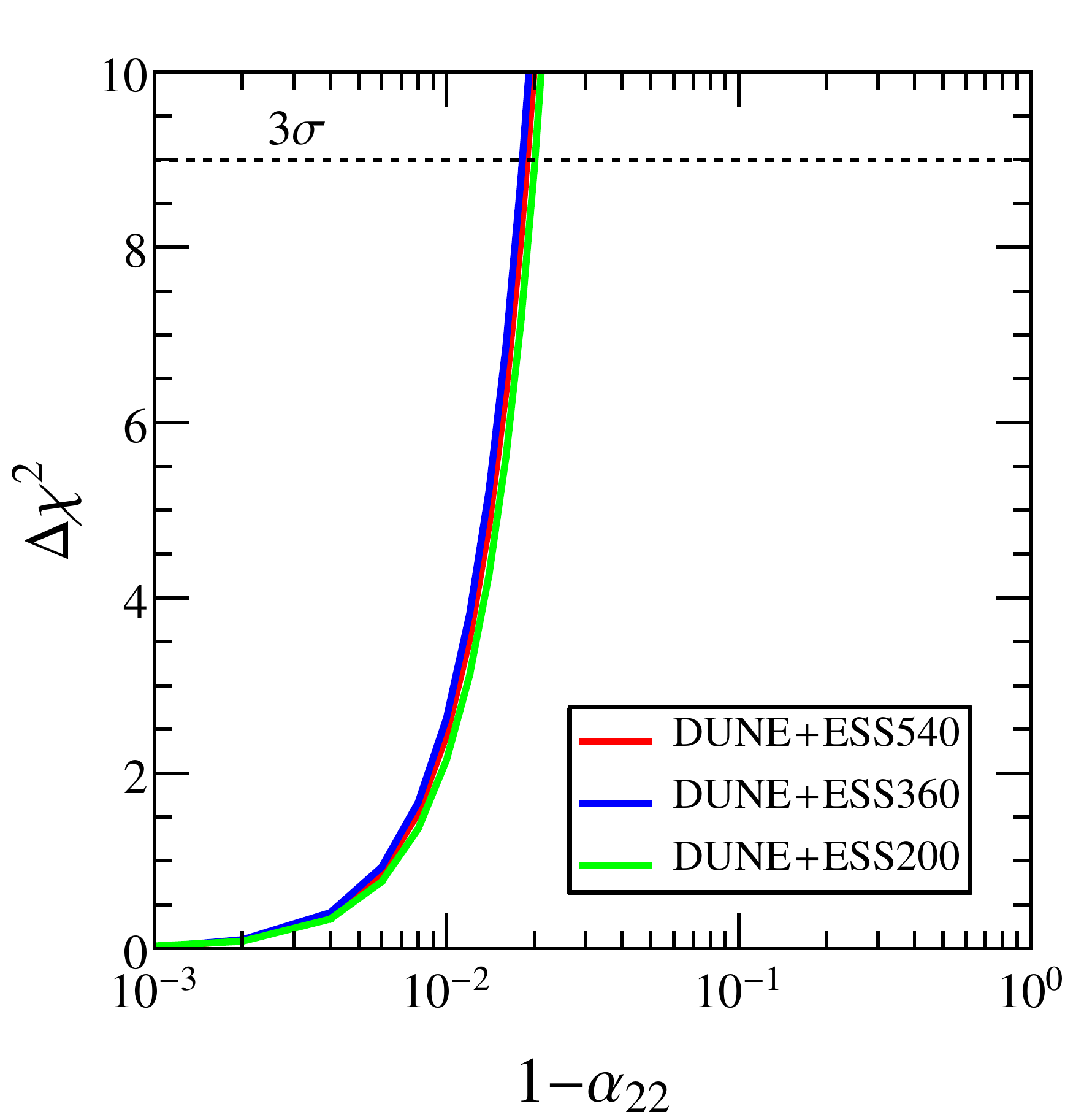}
\caption{
  Upper panels: one dimensional $\Delta\chi^2$ projection on the NU parameters $|\alpha_{21}|$ (left), $\alpha_{11}$ (middle), and $\alpha_{22}$ (right panel) for the three different baselines considered for ESSnuSB (red, blue and green lines, see the legend) and for DUNE (black line). 
Lower panels: combined sensitivities for DUNE and ESSnuSB for the three baselines considered.
 Normal mass ordering has been assumed in the analysis.}
\label{fig:chisq_alpha}
\end{figure*}

To understand better the previous result one can select a point allowed at $1\sigma$, $2\sigma$, and $3\sigma$ for a $540$~km, $360$~km, and $200$~km baseline, respectively.
  We show in Fig.~\ref{events_spectra} the appearance event spectra for the neutrino (upper-left panel) and antineutrino mode (upper-right panel).
  For each of the three different baselines we choose a benchmark point (magenta star in Fig.~\ref{alpha21_phi21}) given by the NU parameter $|\alpha_{21}|=0.04$ and the standard CP phase $\delta_{\rm CP} = -90^\circ$.
  For this benchmark choice, the other NU parameters arising from the $\chi^2$ marginalization take the values shown in Table~\ref{table4}.
 Note that the optimized values of the standard oscillation parameters are not shown here explicitly. 
  We use all of them as input to generate the event spectra in Fig.~\ref{events_spectra}.
  These are shown as dashed lines, while the unitary scenario is represented with solid lines.
  In order to better compare the standard unitary (UN) with the NU case, we show in the lower panels of Fig.~\ref{events_spectra} the absolute differences of the number of events divided by the statistical uncertainty in each bin i.e., $\rm\rm \Delta = |N_{UN}-N_{NU}|/\sqrt{N_{UN}}$. 
This provides a crude measurement of the statistical significance of our unitarity test.
One sees that the best sensitivities to the standard parameter $\delta_{\rm CP}$ and the NU parameter $|\alpha_{21}|$ are achieved for the $200$~km baseline, as already shown in Fig.~\ref{alpha21_phi21}.
Notice also that the $360$~km baseline does somewhat better than the $540$~km baseline.
This is also reflected in the left panels of Fig.~\ref{fig:chisq_alpha}.\\[-.3cm]

In Fig.~\ref{fig:chisq_alpha}, we show the one-dimensional projection
of the $\Delta\chi^2$ as a function of the  NU parameters 
$|\alpha_{21}|$ (left panel), $\alpha_{11}$ (middle panel), and $\alpha_{22}$ (right
panel). 
The red, blue, and green curves in the upper panels show the
sensitivities corresponding to the three ESSnuSB baselines, 540 km,
360 km, and 200 km, respectively, while the black curve corresponds to
the DUNE sensitivity. In the lower panels, the same color code is used to
represent the sensitivities for the combined analysis of DUNE and the ESSnuSB for baselines of 540, 360, and 200 km.

As before, this figure is obtained assuming the standard unitary framework as the true hypothesis, and testing the NU framework against it, i.e. $\Delta\chi^2 = \chi^2_{\rm NU} - \chi^2_{\rm UN}$.
The true values of the oscillation parameters are taken from Table~\ref{table1}, fixing the two solar parameters $\theta_{12}$ and $\Delta m_{21}^2$ at their benchmark choices,
and marginalizing over $\theta_{13}$, $\theta_{23}$, and $\Delta m_{31}^2$. 
Since the exact value of the standard CP phase $\delta_{\rm CP}$ is currently not accurately known, we have marginalized over its true and test values within its full range. 

For the sensitivity to $|\alpha_{21}|$, shown in the left panels of Fig.~\ref{fig:chisq_alpha}, we have freely varied the NU parameters $\alpha_{11}$, $\alpha_{22}$, and $\alpha_{33}$
  within their allowed $3\sigma$ ranges as given in Table~\ref{table3} and the phase $\phi_{21}$ from $-\pi$ to $+\pi$.
The other non-diagonal NU-parameters $|\alpha_{31}|$ and $|\alpha_{32}|$ have been set  to zero. 
From the upper left panel we can see that the best sensitivity for the ESSnuSB comes from the 200 km baseline, followed by the 360 km and 540 km baselines. 
The expected $3\sigma$ upper limits on $|\alpha_{21}|$ corresponding to 540 km, 360 km, and 200 km baselines are 0.16, 0.075, and 0.04, respectively. 
These limits would be independent and complementary to those given in Table~\ref{table3} and illustrate the ESSnuSB potential in probing new physics. 
For the case of DUNE, the upper limit at $3\sigma$ would be 0.046.
One sees the sensitivities expected at ESSnuSB are competitive and, for the smaller baseline, it performs slightly better than DUNE~\cite{Escrihuela:2016ube}.
In the lower left panel of Fig.~\ref{fig:chisq_alpha}, we show how the combined sensitivities of DUNE and ESSnuSB, improve with respect to the individual sensitivities.
The $3\sigma$ upper limits corresponding to the DUNE+ESS540, DUNE+ESS360, and DUNE+ESS200 combinations would be 0.04, 0.03, and 0.022, respectively.
  It is encouraging to see that the combination of DUNE and the 200 km baseline of ESSnuSB would be able to set a constraint on $|\alpha_{21}|$ which is better than the current $3\sigma$ upper bound given in Table~\ref{table3}.

The middle panel of Fig.~\ref{fig:chisq_alpha} shows the sensitivities to the diagonal NU parameter $\alpha_{11}$. 
The analysis method is very similar to the previous one as far as the standard unitarity-parameters are concerned.
However, for the NU parameters, we marginalize over the parameters $\alpha_{22}$,
$|\alpha_{21}|$, and its associated new CP-phase $\phi_{21}$.
As before, the off-diagonal parameters $\alpha_{31}$ and $\alpha_{32}$ have been set to zero and $\alpha_{33}$ was set to unity.
  One can see that the $3\sigma$ sensitivities on $\alpha_{11}$ corresponding to the 540, 360, and 200 km baselines are lower than for DUNE, i.e. $0.87$ versus $0.95$.
From the lower middle panel one sees that the combined sensitivities corresponding to DUNE+ESS540, DUNE+ESS360, and DUNE+ESS200 are very similar and
approximately equal to $0.955$, which would somewhat improve the current sensitivity on $\alpha_{11}$.

The sensitivities on $\alpha_{22}$ are shown in the right panels of Fig.~\ref{fig:chisq_alpha}. In this case, we follow exactly the same steps as in the $\alpha_{11}$ analysis but marginalizing over $\alpha_{11}$.
Our results show that the best performance is expected from DUNE, followed by the ESSnuSB with baselines 360 km, 540 km, and 200 km.
The expected sensitivities for the individual setups do not improve over the current lower bound.
However, one sees that the combined sensitivity of DUNE and any ESSnuSB baseline gives a slightly lower sensitivity compared to the current $3\sigma$ lower bound.
One concludes that all these sensitivities, specially those coming from the combined analyses, are encouraging and will play an important role especially when combined with the already available data.
For the sake of convenience we quote all our bounds as Table~\ref{table5}.
Note also that these NU parameter sensitivities at the ESSnuSB may be complemented by additional information coming from future experiments such as T2HK or JUNO~\cite{Dutta:2019hmb, JUNO:2015zny}. 
\begin{table}[h!]
{%
\begin{center}
\begin{tabular}{|c|c|c|c|}
\hline
 3$\sigma$ sensitivity &\hspace{0.3cm} $|\alpha_{21}|$ \hspace{0.3cm} &\hspace{0.3cm}  $\alpha_{11}$ \hspace{0.3cm} &\hspace{0.3cm} $\alpha_{22}$\hspace{0.3cm}  \\ 
\hline
ESS540   & $< 0.160$ & $> 0.850$ & $ > 0.964$   \\
\hline
ESS360   & $< 0.075$ & $ >0.860$ & $ > 0.973$   \\
\hline
ESS200   & $< 0.040$ & $ >0.870$ & $ > 0.967$   \\
\hline
DUNE     & $< 0.046$ & $ >0.945$ & $ > 0.975$   \\
\hline
DUNE+ESS540    & $< 0.040$ & $> 0.951$ & $ > 0.981$   \\
\hline
DUNE+ESS360    & $< 0.030$ & $> 0.952$ & $ > 0.982$   \\
\hline
DUNE+ESS200    & $< 0.022$ & $ >0.954$ & $ > 0.979$   \\
\hline
\end{tabular}
\end{center}
}%
\caption{ESSnuSB, DUNE, and  combined $3\sigma$ sensitivities on the NU parameters $|\alpha_{21}|$, $\alpha_{11}$, and $\alpha_{22}$ obtained in this work.}
\label{table5}
\end{table}

\subsection{CP violation discovery potential} 
\label{sec:cp-viol-disc}

\begin{figure*}[h!]
\centering
\includegraphics[height=6.9cm,width=6.9cm]{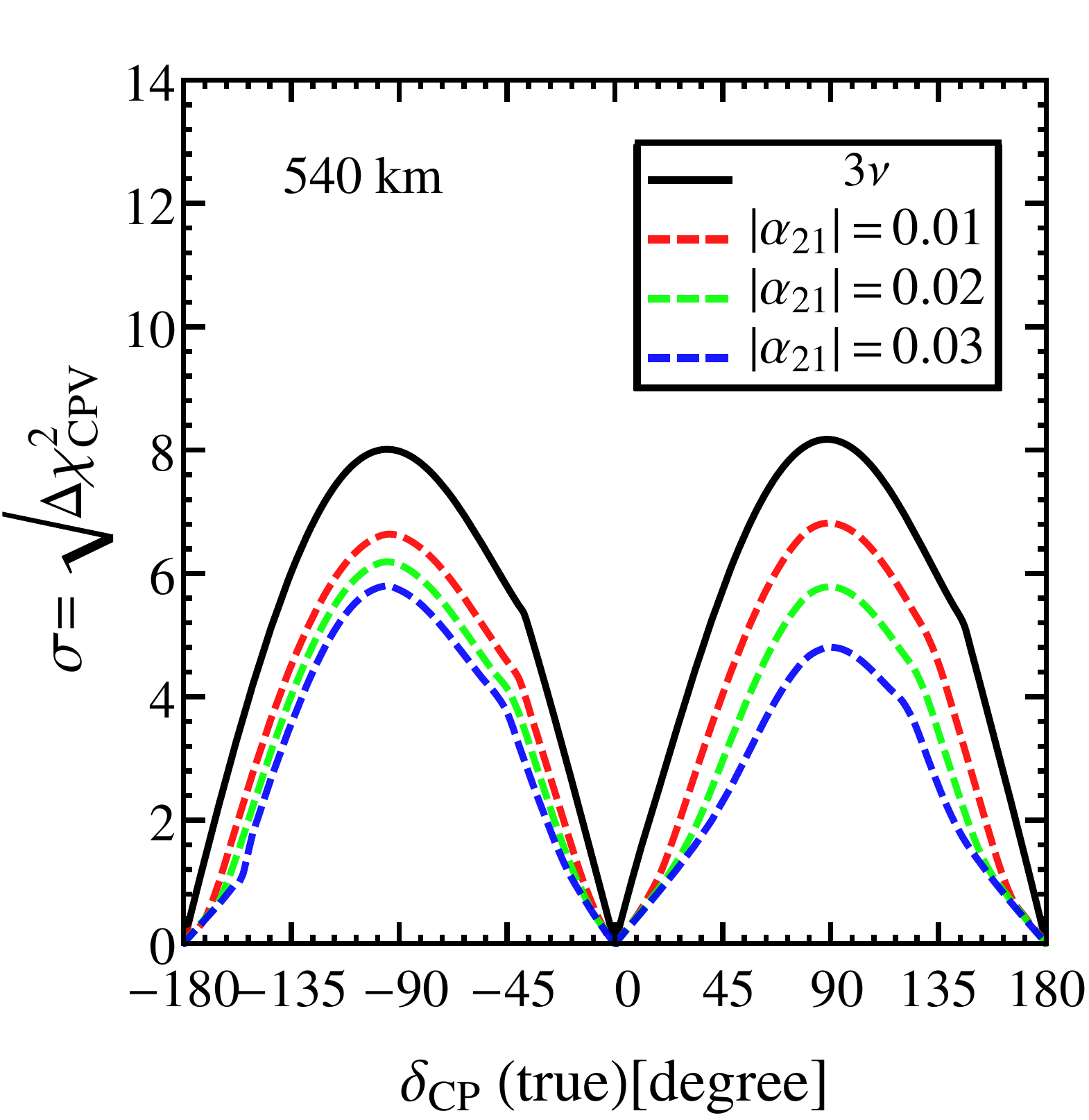}
\includegraphics[height=6.9cm,width=6.9cm]{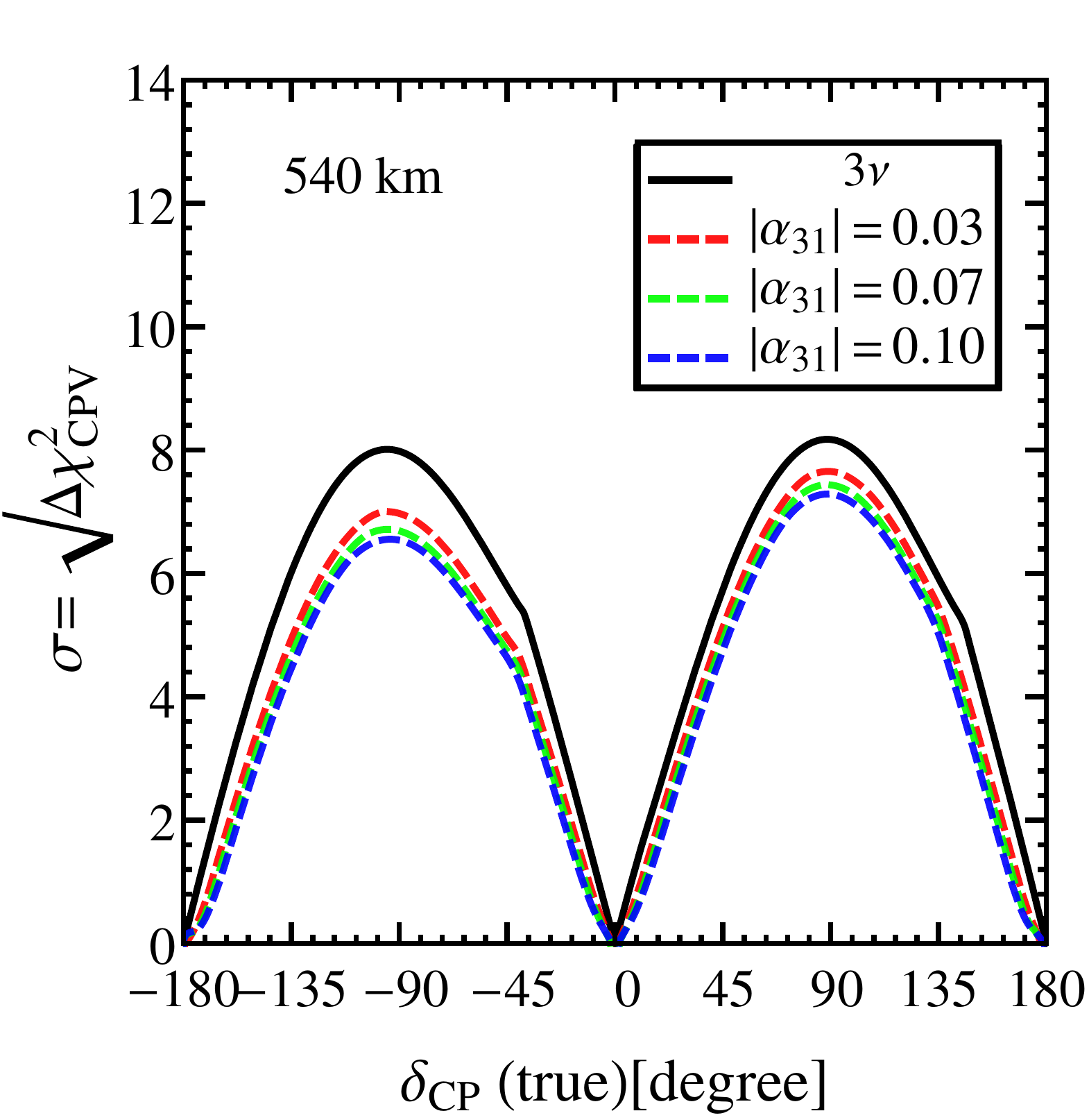}\\
\includegraphics[height=6.9cm,width=6.9cm]{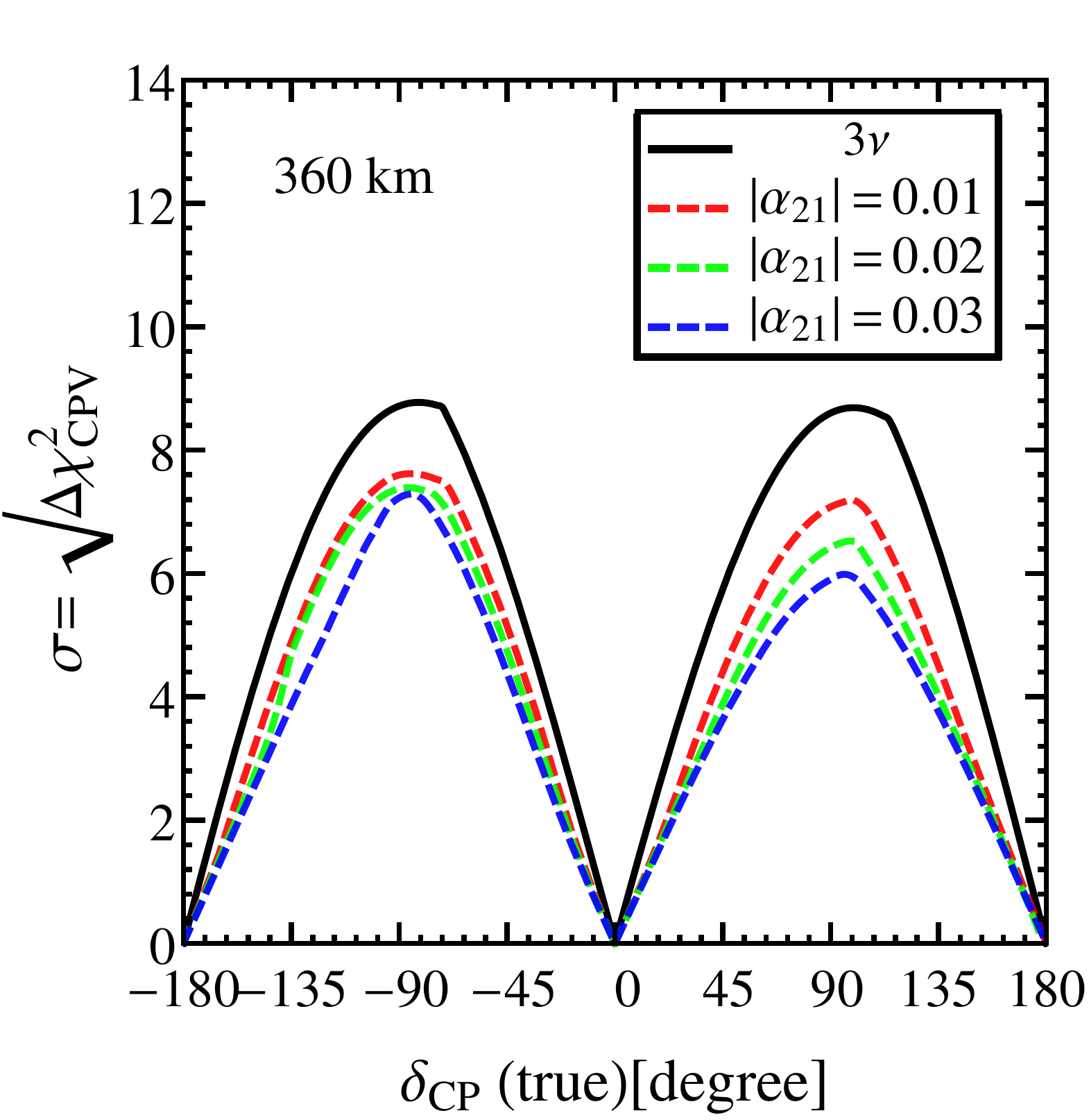}
\includegraphics[height=6.9cm,width=6.9cm]{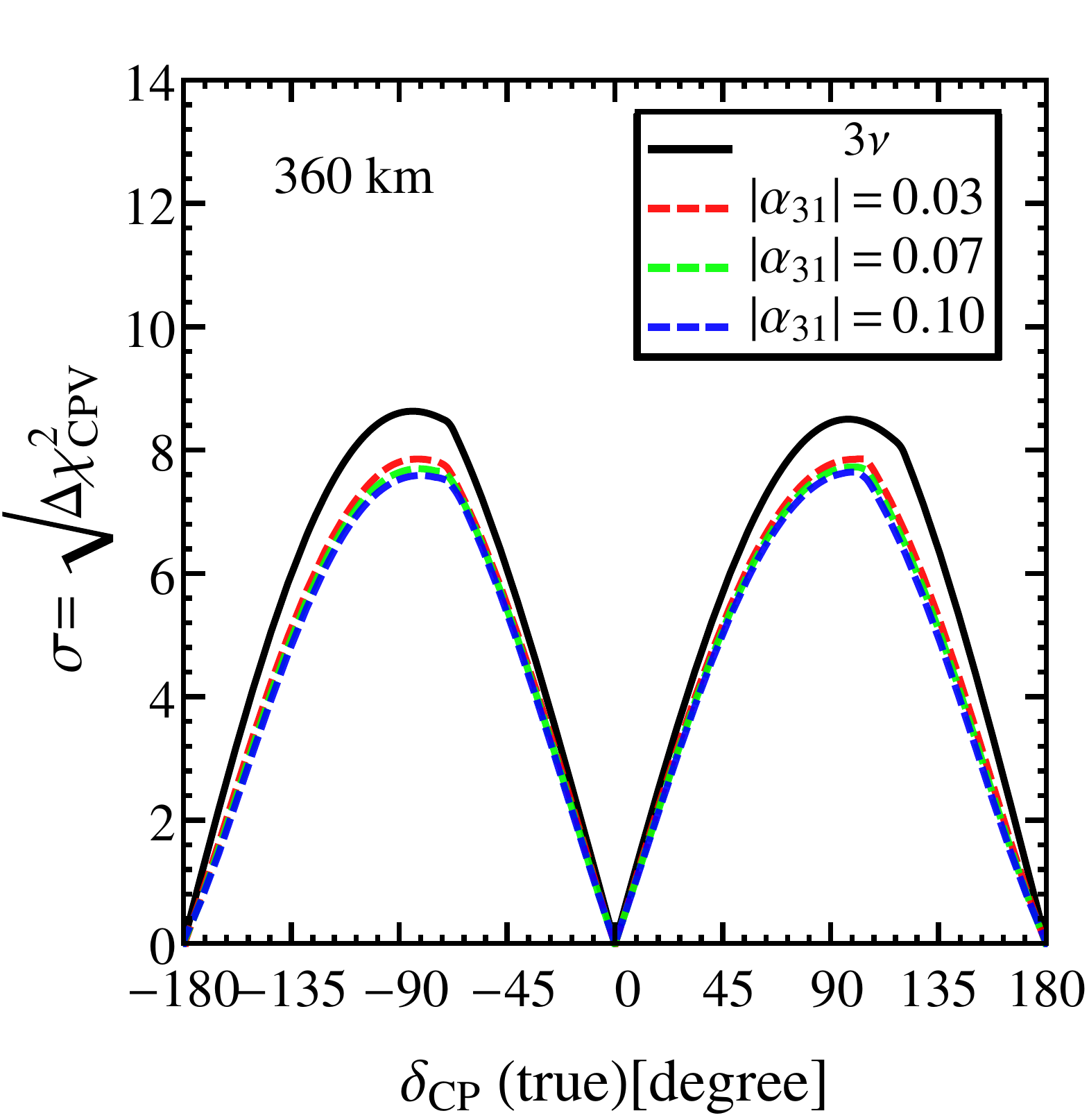}\\
\includegraphics[height=6.9cm,width=6.9cm]{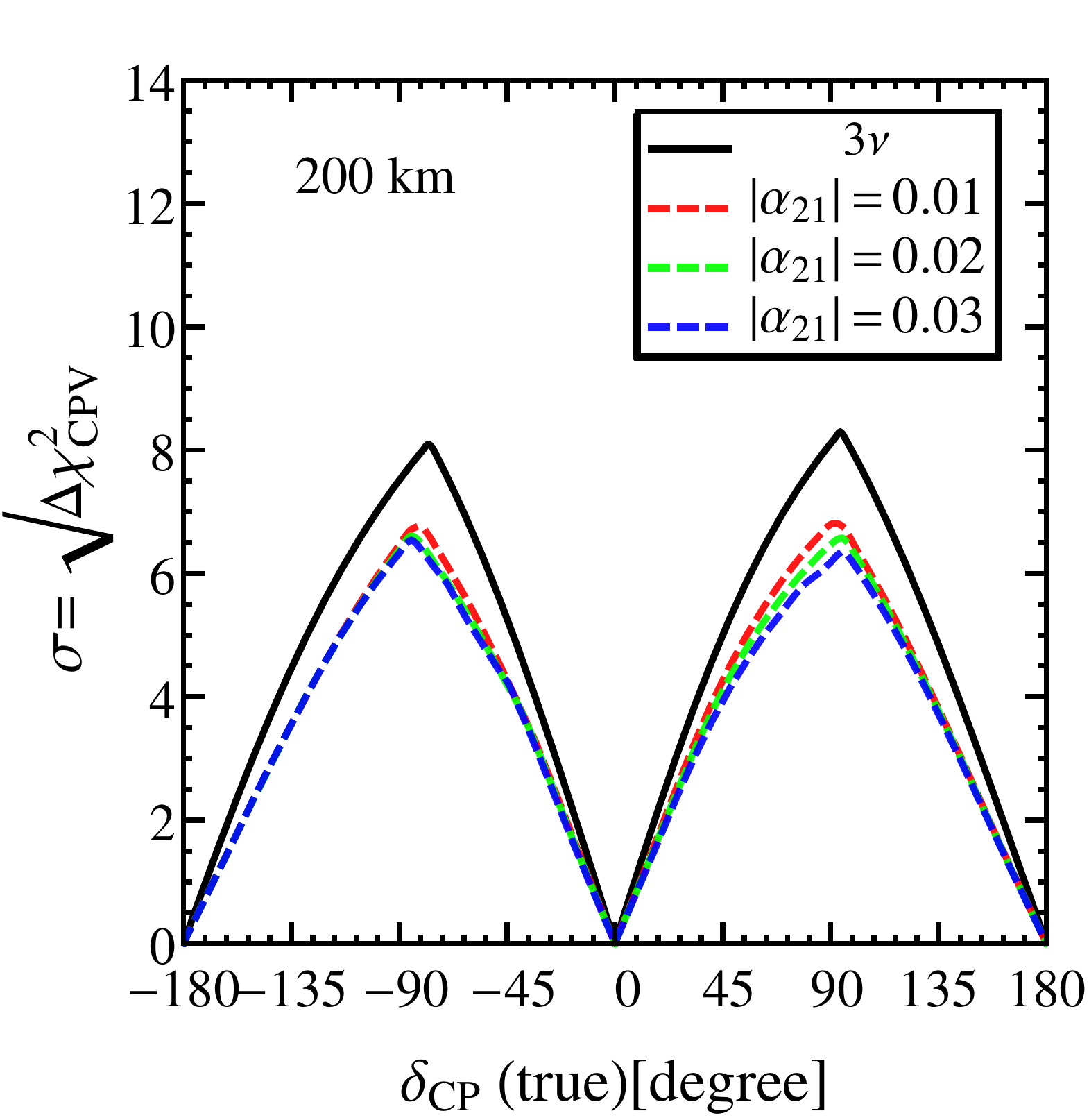}
\includegraphics[height=6.9cm,width=6.9cm]{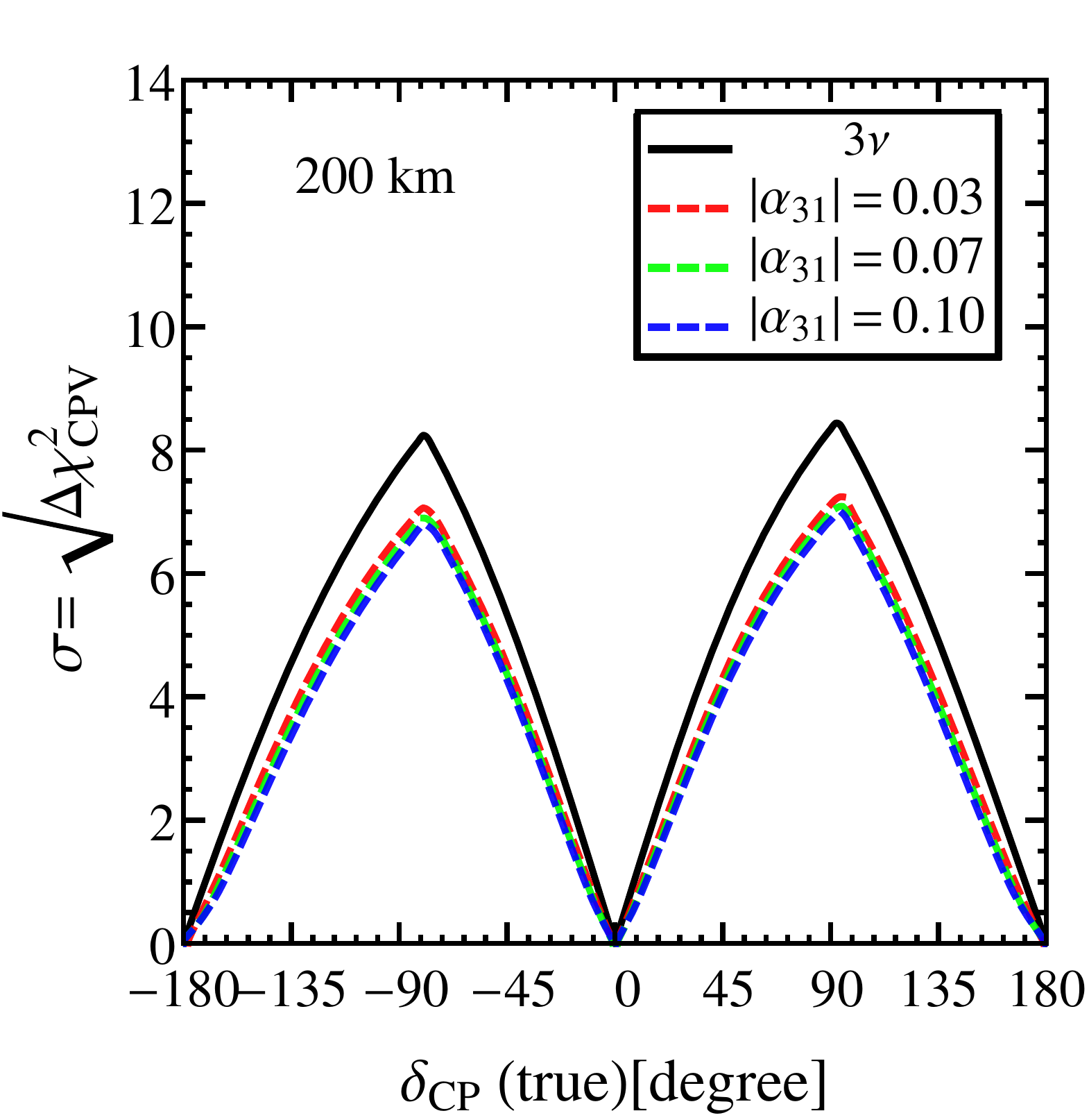}
\caption{CPV sensitivities at ESSnuSB for various baselines, as indicated. The solid black lines correspond to unitary neutrino mixing, while dashed lines show the sensitivities in the presence of unitarity violation. The value of $|\alpha_{21}|$ ($|\alpha_{31}|$) has been fixed in the left (right) panel  in the data as well as in the theory. We have marginalized over $\theta_{13}$, $\theta_{23}$, $\Delta m_{31}^2$, as well as over the  true and test values of $\alpha_{11}$, $\alpha_{22}$, $\alpha_{33}$, and $\phi_{21}$ (in the left panel) and $\phi_{31}$ (in the right panel), within their allowed ranges.}
\label{fig:cpv}
\end{figure*}

\begin{figure*}[t!]
\centering
\includegraphics[height=6.9cm,width=6.9cm]{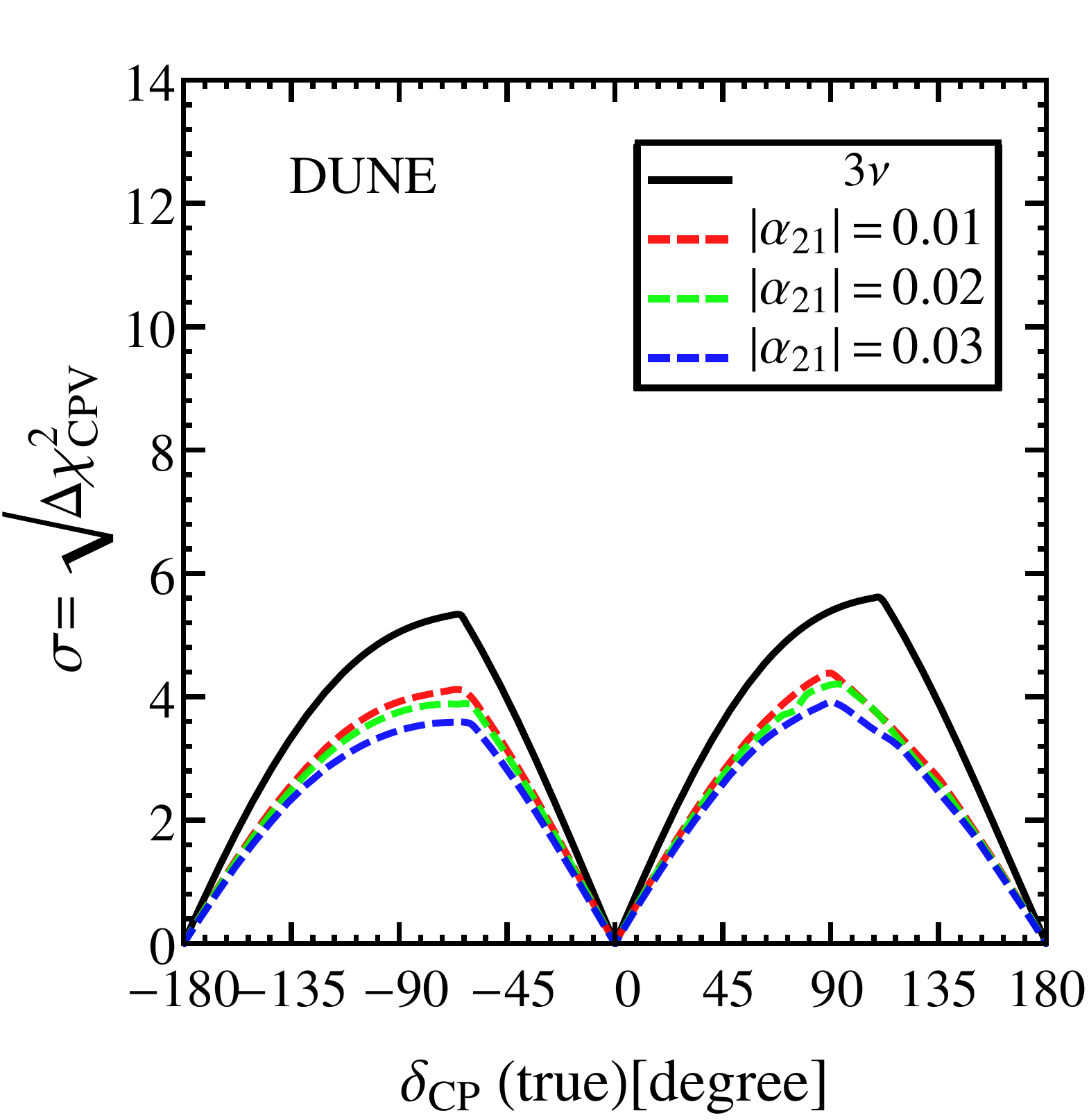}
\includegraphics[height=6.9cm,width=6.9cm]{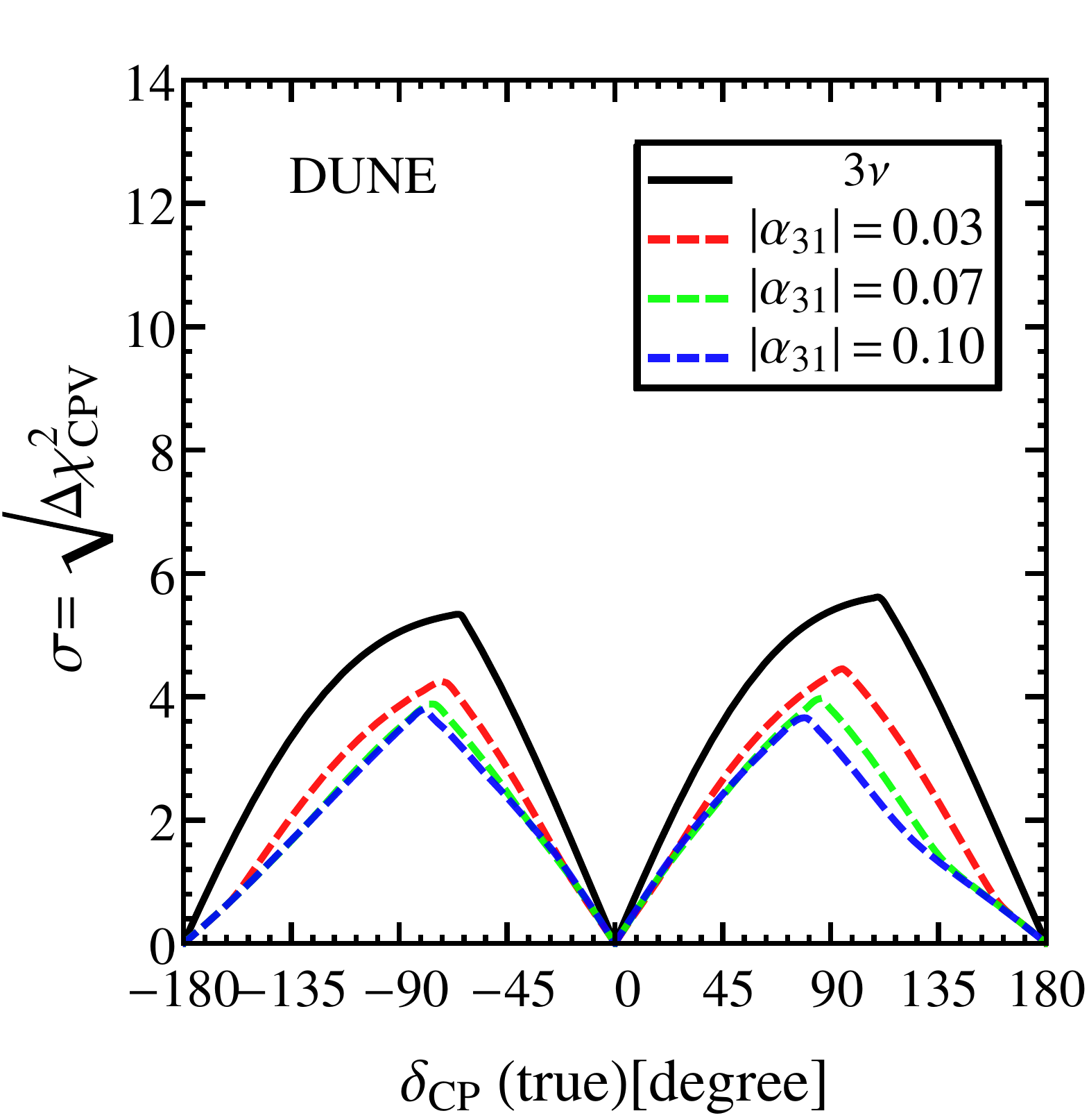}
\caption{CPV sensitivities at DUNE. Conventions and analysis procedure are the same as in Fig.~\ref{fig:cpv}, see its caption for details.}
\label{fig:cpv2}
\end{figure*}

\begin{figure*}[t!]
\centering
\includegraphics[height=6.9cm,width=6.9cm]{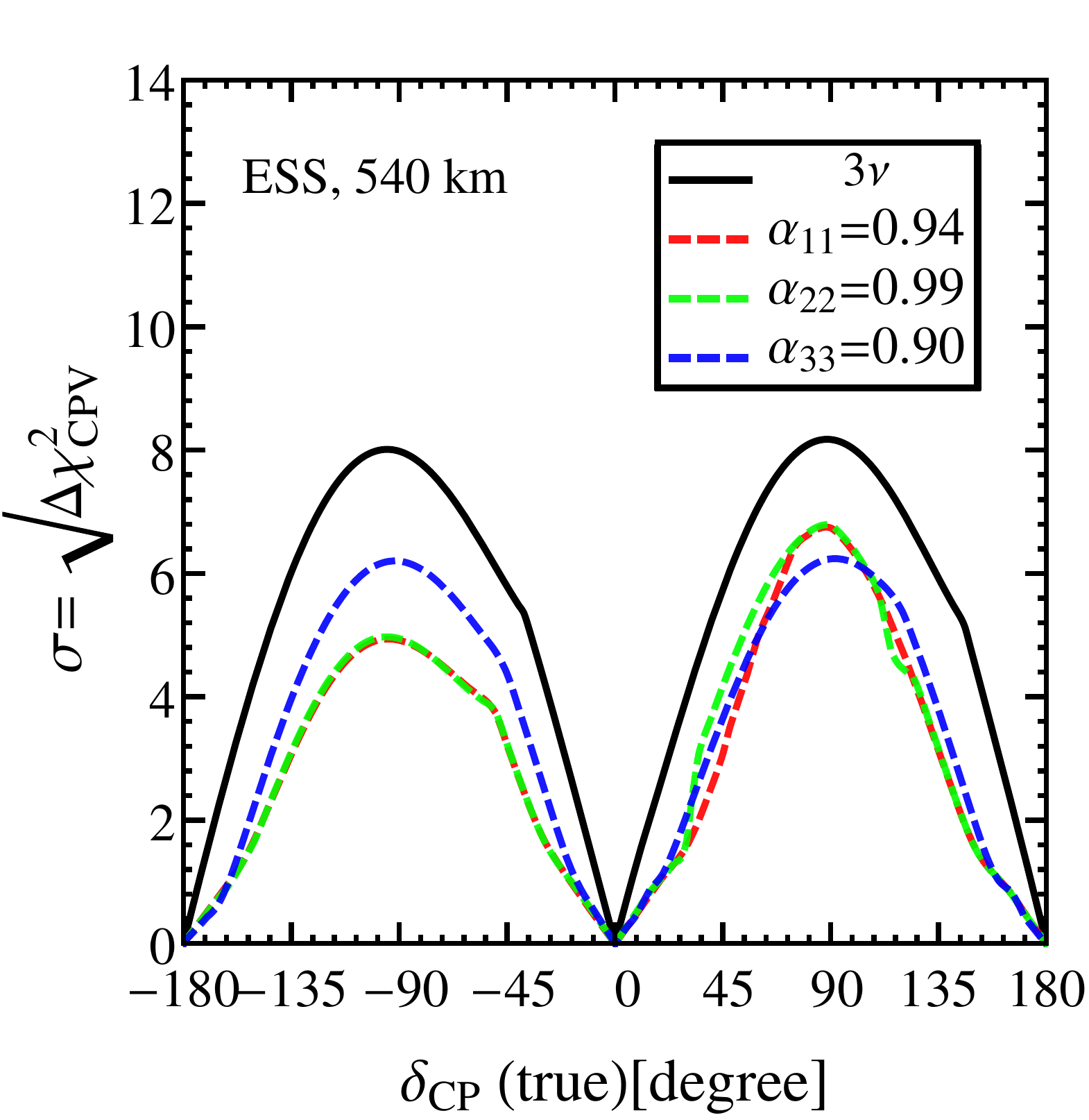}
\includegraphics[height=6.9cm,width=6.9cm]{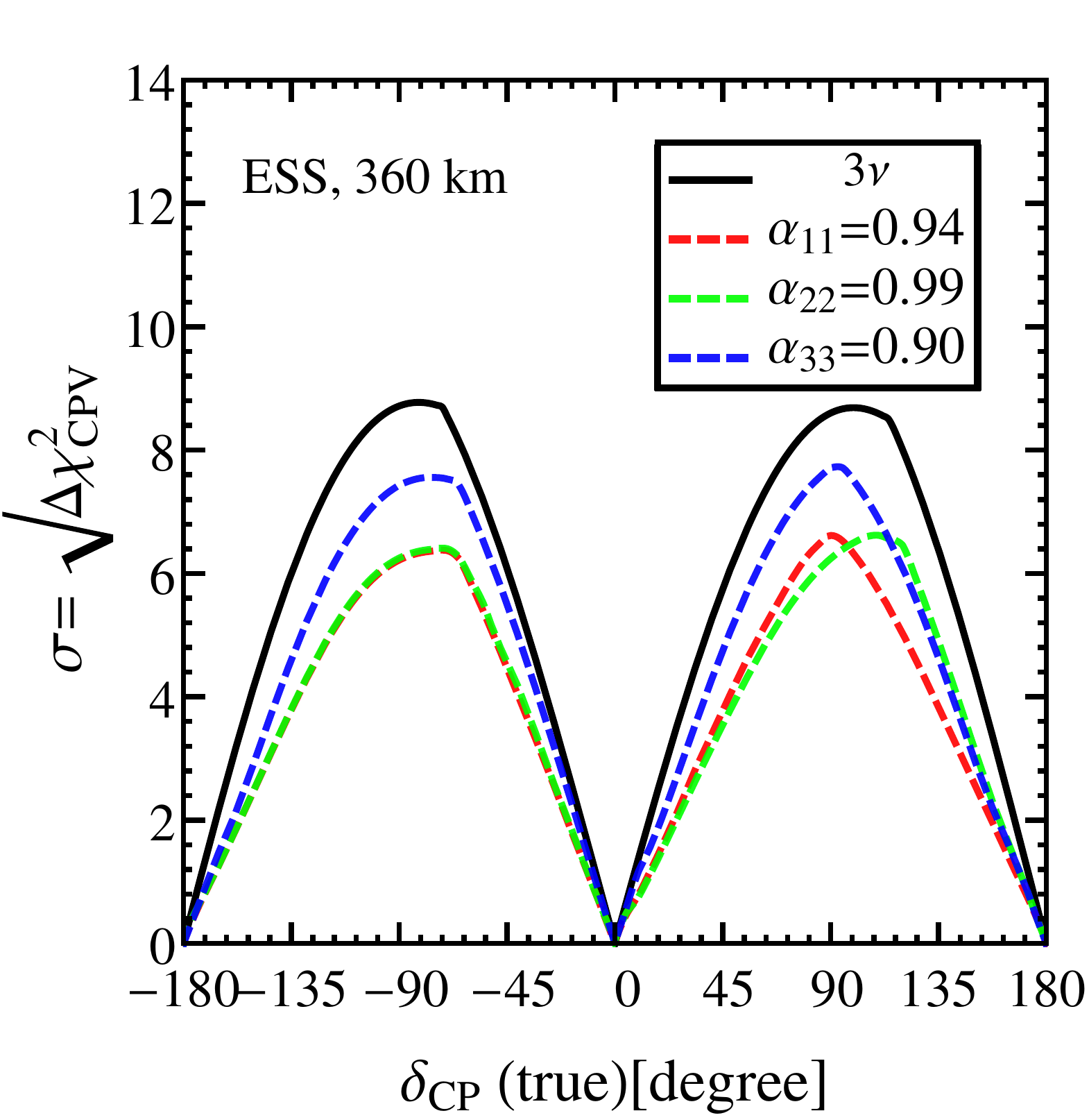}
\includegraphics[height=6.9cm,width=6.9cm]{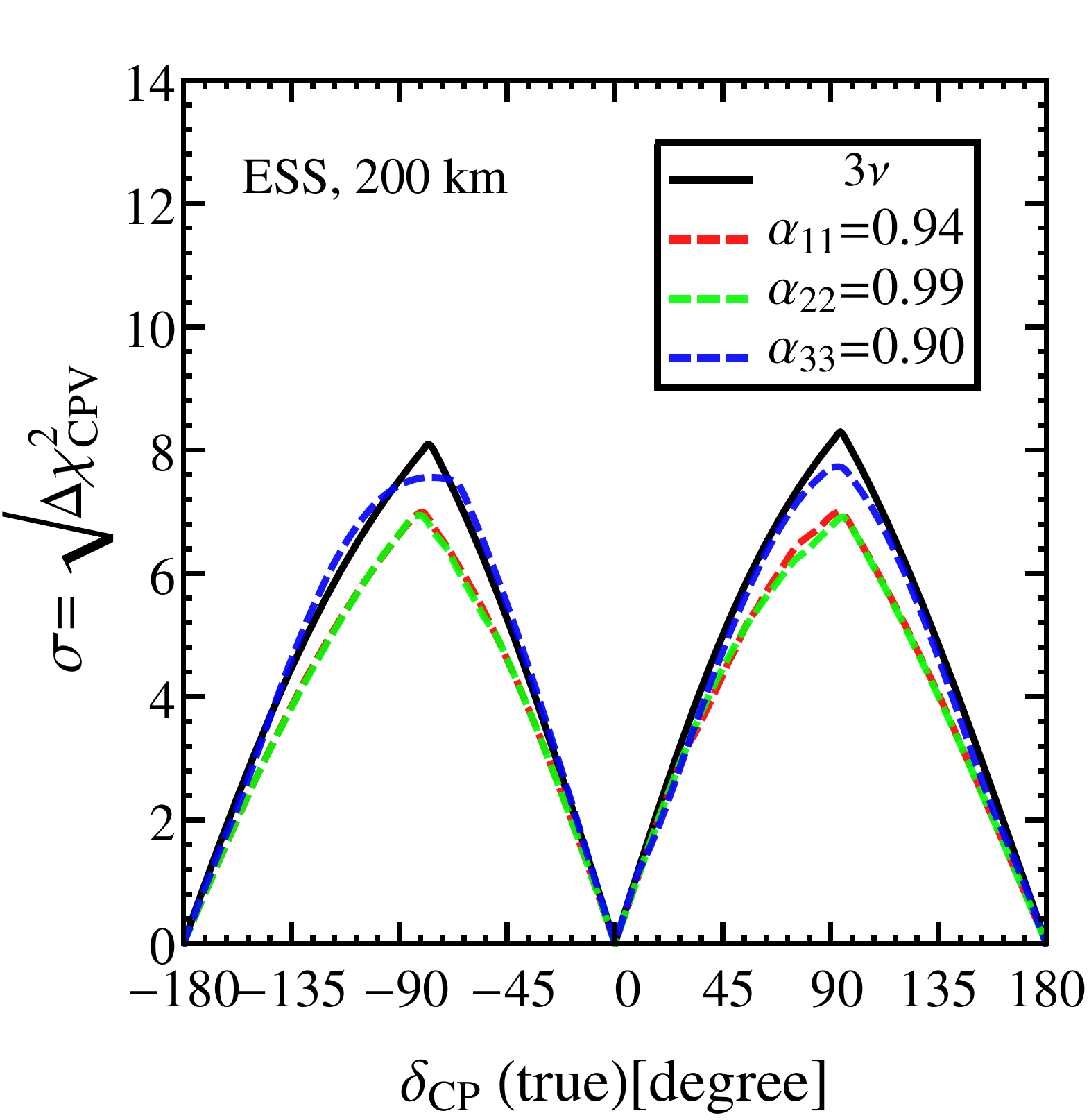}
\includegraphics[height=6.9cm,width=6.9cm]{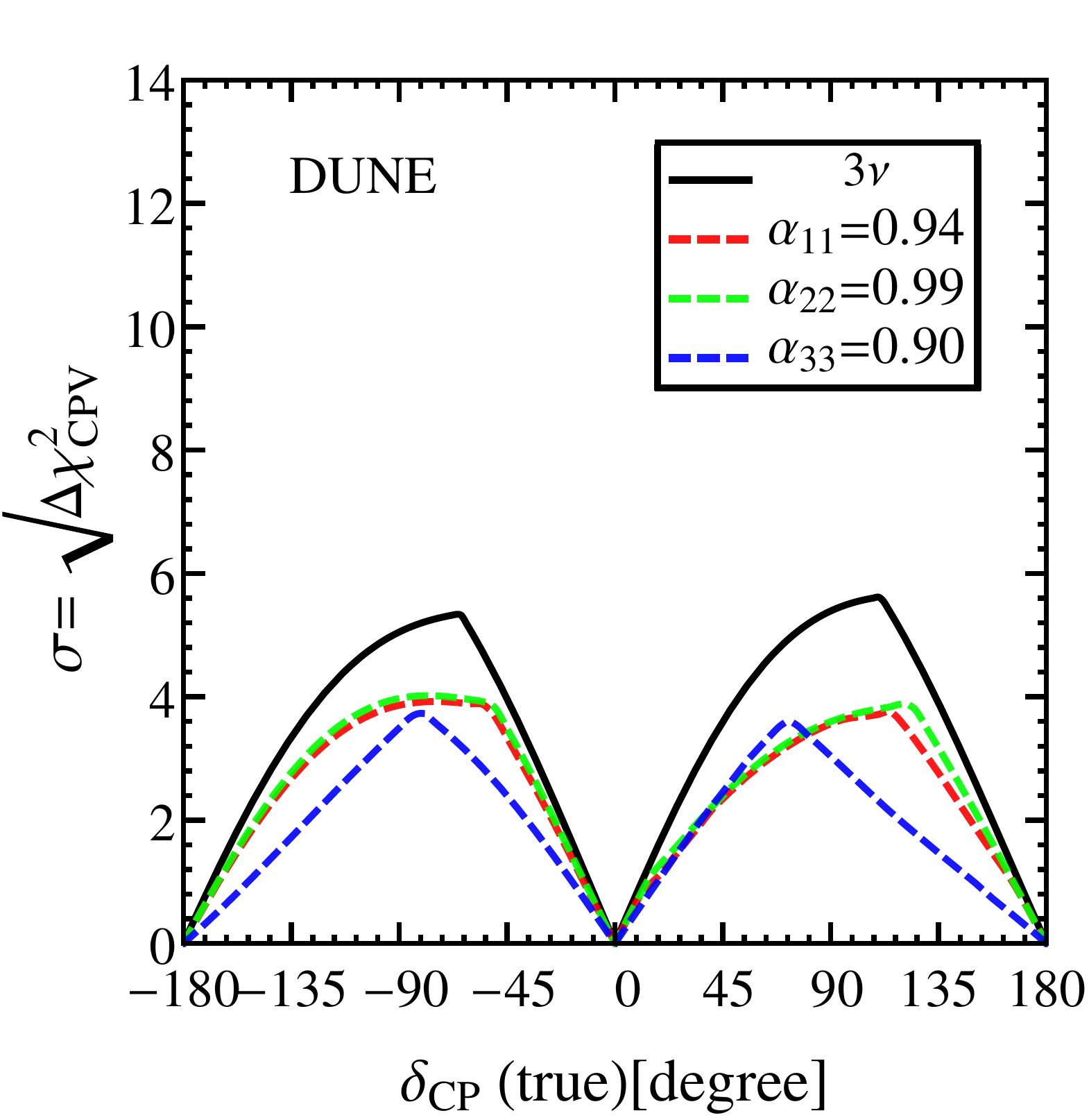}
\caption{
  CPV sensitivities at ESSnuSB for the three baselines under consideration and for DUNE. 
   The red, green, and blue dashed lines in each panel correspond to the CPV sensitivities for the fixed values of $\alpha_{11}=0.94$, $\alpha_{22}=0.99$, and $\alpha_{33}=0.90$, respectively, in the data as well as in the theory. For simulation details, see the text.}
\label{fig:cpv3}
\end{figure*}

Now we turn back to the ``conventional'' CP violation discovery potential within our generalized nonunitary framework.
  The CP violation (CPV) discovery potential of the ESSnuSB and DUNE setup 
  is summarized in Figs.~\ref{fig:cpv}, \ref{fig:cpv2} and \ref{fig:cpv3}.
Our results are given both for the unitary and  the nonunitary  framework.
As in the standard $\delta_{\rm CP}$ sensitivity study~\cite{DUNE:2015lol}, the CP-violating hypothesis is tested against a CP-conserving scenario through~\cite{Escrihuela:2016ube}
\begin{align}
&\Delta\chi^2 (\delta_{\rm CP}^{\rm true}) = \nonumber \\
&\rm{Min}\left[\Delta\chi^2\left(\delta_{\rm CP}^{\rm true}, \delta_{\rm CP}^{\rm test} = 0 \right),\,\,\Delta\chi^2\left( \delta_{\rm CP}^{\rm true}, \delta_{\rm CP}^{\rm test} = \pm \pi\right)\right].
\end{align}
This way, we obtain the significance with which one can reject the
test hypothesis of no CP violation.
We have assumed five nonzero NU parameters: the three diagonal ones (marginalized over the true and test values), plus one non-diagonal parameter, either $|\alpha_{21}|$ (left panels)
or $|\alpha_{31}|$ (right panels), with the associated complex CP
phase ($\phi_{21}$ or $\phi_{31}$), also marginalized.
Concerning the standard three-neutrino parameters (both in the unitary and nonunitary scenarios), we have marginalized over the two mixing angles
  $\theta_{13}$ and $\theta_{23}$ and the ``atmospheric'' mass-squared splitting $\Delta m_{31}^2$.
The left (right) column of Fig.~\ref{fig:cpv} represents the CPV discovery potential of ESSnuSB for the three baseline choices and
three different values of the NU parameter $|\alpha_{21}|$ ($|\alpha_{31}|$).
The black solid line in each plot corresponds to the standard three-neutrino unitary framework.
Upper, middle, and lower panels represent the results obtained for 540, 360, and 200 km baselines, respectively.
The red, green, and blue dashed curves in the left plots represent the CPV discovery sensitivities in the NU framework corresponding to different values
 of $|\alpha_{21}|$, $0.01,\, 0.02,\, \rm{and}\, 0.03$, respectively (the latter, relatively large value, is taken mainly for comparison).
One sees that the CPV sensitivity in the standard unitary neutrino oscillation framework always lies around
  $8\sigma$ for $\delta_{\rm CP}(\rm true) = \pm 90^\circ$ for all three baselines, see the solid black line in all the panels.
This agrees with the results presented in Refs.~\cite{Baussan:2013zcy,Agarwalla:2014tpa,Alekou:2021coj}. 
All baselines have more or less similar sensitivities, except for the fact that the $\delta_{\rm CP}$ range over which CPV can be established for 540 km and 360 km is slightly bigger
than for 200 km.
This fact is also confirmed in Ref.~\cite{Baussan:2013zcy}. However, we will see the merits of the 200 km baseline in what follows. 
As far as the  NU framework is concerned,  two of our benchmark values, $|\alpha_{21}| = 0.01,\, \rm{and}\,0.02$ lie within the current $3\sigma$ limit,
whereas $|\alpha_{21}| = 0.03$ lies slightly outside the current allowed limit. 
\footnote{As already mentioned, the line with 0.03 is included mainly for comparison, to have a broader view of degrading behaviour
  with respect to the non-diagonal NU parameters.}.\\[-.3cm]

We stress that the CPV discovery sensitivity is degraded with respect to the unitary case.
  This is expected, due to the presence of new phases associated to unitarity violation~\cite{Miranda:2016wdr}.
Clearly, the CPV discovery sensitivity decreases with the increasing values of $|\alpha_{21}|$, specially for 540 km,
leading to a minimum  sensitivity of $4.6\sigma$ for $\delta_{\rm CP}(\rm true) = \pm 90^\circ$.  
The deterioration of the CPV sensitivity is smaller for 360 km and, with a minimum  sensitivity of around $5.7\sigma$ for $\delta_{\rm CP}(\rm true) = \pm 90^\circ$.  
For the 200 km baseline, the deterioration further reduces, with a minimum sensitivity of $6.1\sigma$  for all three benchmark choices, at $\delta_{\rm CP}(\rm true) = \pm 90^\circ$.
All in all, one sees that the degrading in CP sensitivity is not as large as one might expect, showing the robustness of the oscillation picture with respect to unitarity violation.
The best sensitivities to $\delta_{\rm CP}$ and the nonunitary parameter $|\alpha_{21}|$ are achieved for a 200 km baseline.\\[-.3cm]

We now turn to the right panels of Fig.~\ref{fig:cpv}. There, we repeated the same analysis for the non-diagonal parameter $|\alpha_{31}|$ and its associated CP phase, $\phi_{31}$.
The three benchmark choices considered for $|\alpha_{31}|$ are $0.03,\, 0.07,\,\rm{and},\,0.10$\footnote{Note that, as in the previous case, the third value of  $|\alpha_{31}|$ lies slightly outside the 3$\sigma$ current limit, and is taken only for comparison.}.
In this case, one finds a mild deterioration of the CPV sensitivity in comparison to the unitary framework for all baselines,
so the impact of $|\alpha_{31}|$ is not significant for 540 km, and negligible for 360 km and 200 km baselines.
 This is due to the fact that $|\alpha_{31}|$ does not appear in the vacuum appearance probability, Eq.~(\ref{pme_vac}), and also
 because of the lower matter effects for ESSnuSB with respect to DUNE~\cite{Escrihuela:2016ube,DUNE:2020fgq}.
As expected, one finds a negligible impact on the CPV sensitivity arising from unitarity violation in this case. Note, however that this will not be the case for DUNE and, as a result, the combined analysis with ESSnuSB will be useful to recover the maximum CPV sensitivity in this case.

Next, in Fig.~\ref{fig:cpv2},  we show the CPV  sensitivities for DUNE. As before, the black solid lines in each panel correspond to the CPV sensitivity in the standard
  unitary framework, showing that DUNE can establish CPV discovery above $5\sigma$ for $\delta_{\rm CP}\,(\rm true)=\pm90^\circ$.
  The red, green, and blue dashed curves in the left panel represent the CPV discovery sensitivities in the NU framework corresponding to different values of $|\alpha_{21}|$:
  $0.01,\, 0.02,\, \rm{and}\, 0.03$, respectively. Similarly, in the right panel, these lines correspond to the CPV sensitivities for $|\alpha_{31}|$ equal to  $0.03,\, 0.07,\, \rm{and}\, 0.10$.
  The simulation method followed in both the panels is the same as for Fig.~\ref{fig:cpv}.
  From the left panel, we see that the CPV sensitivity in the presence of NU is degraded with respect to the unitary framework, according to the value of $|\alpha_{21}|$.
  As a result, for the maximal values of $\delta_{\rm CP}$(true) one can only ensure a minimum CPV sensitivity of $3.6\sigma$.
  Similarly, from the right panel, we can see that the CPV sensitivity decreases with increasing $|\alpha_{31}|$ with a minimum sensitivity of $3.6\sigma$
  for $\delta_{\rm CP}\,(\rm true)=\pm 90^\circ$.

Comparing Figs.~\ref{fig:cpv} and \ref{fig:cpv2} one sees that, for the conventional unitary scenario the CPV sensitivity of ESSnuSB is
    better than that of DUNE. Also the deterioration due to NU effects is smaller than for DUNE,  where the matter effect is large.
  Hence, in this case, ESSnuSB will play a complementary role to DUNE in achieving a robust CP violation sensitivity in the presence of nonunitarity.

  So far we have analyzed the effect of the non-diagonal NU parameters on the CPV sensitivity of the experimental setups under consideration.
  Now we will investigate the role of the diagonal nonunitary parameters, $\alpha_{11}$, $\alpha_{22}$, and $\alpha_{33}$.
  The corresponding CPV sensitivities in this case for ESSnuSB and DUNE are shown in the four panels of Fig.~\ref{fig:cpv3}.
The results for the standard unitary case (solid black line) for both the ESSnuSB and DUNE have been already described in Fig.~\ref{fig:cpv} and Fig.~\ref{fig:cpv2}. 
The red, green, and blue dashed lines in each panel correspond to the CPV sensitivities for  fixed values of $\alpha_{11}=0.94$, $\alpha_{22}=0.99$, and $\alpha_{33}=0.90$,
respectively, in the data as well as in the theory. 
The procedure we have followed for this analysis is slightly different from what we have used for the non-diagonal parameters.
In order to evaluate the CPV sensitivity in the NU framework for $\alpha_{11}=0.94$, we have marginalized over
$\alpha_{22}$, $|\alpha_{21}|$, and $\phi_{21}$  within their allowed $3\sigma$ ranges.
All other off-diagonal parameters are assumed to be zero, and we have also fixed $\alpha_{33}=1$ in the data as well as in the theory.
  The analysis method for $\alpha_{22}$ is analogous to the previous one, where we have marginalized over $\alpha_{11}$, $|\alpha_{21}|$, and $\phi_{21}$.
    On the other hand for $\alpha_{33}$, we marginalized over $\alpha_{11}$, $|\alpha_{31}|$, and $\phi_{31}$, fixing $\alpha_{22}=1$.

One sees that ESSnuSB performs better than DUNE for the overall CPV discovery sensitivity. 
  Of course, in all cases these sensitivities are degraded substantially with respect to those of the unitary case. 
In fact, the CPV discovery for DUNE falls below $4\sigma$ for any value of $\delta_{\rm CP}\,(\rm true)$.
As seen in Fig.~\ref{fig:cpv3} the best result is obtained for the 200 km baseline, followed by 360 km of ESSnuSB.

In short, the ESSnuSB will play a crucial role in securing a robust CPV discovery sensitivity in the presence of unitarity deviations close to the current upper bound.

\subsection{CP reconstruction}
\label{sec:cp-reconstruction}

In our simplest scenario there are two relevant CP phases, the standard three-neutrino Dirac phase $\delta_{\rm CP}$ and the phase $\phi_{21}$ associated to
nonunitarity~\footnote{Note that $|\alpha_{31}|$ and $|\alpha_{32}|$ enter only in the appearance neutrino probability through matter effects,
  strongly suppressing the sensitivity to the associated phases $\phi_{31}$ and $\phi_{32}$.}.
One can therefore have four ``CP conserving'' cases, when either of them equals 0 or $\pi$.
Likewise, four cases in which one has ``maximal'' CP violation, defined by having the modulus of any of them equal to $\pi/2$.
  In this section, we discuss how well the European spallation source setups can reconstruct the standard CP phase $\delta_{\rm CP}$ as well as the nonunitarity phase $\phi_{21}$ for a few selected benchmarks. 
  
\begin{figure*}[t!]
\centering
\includegraphics[width=.95\textwidth]{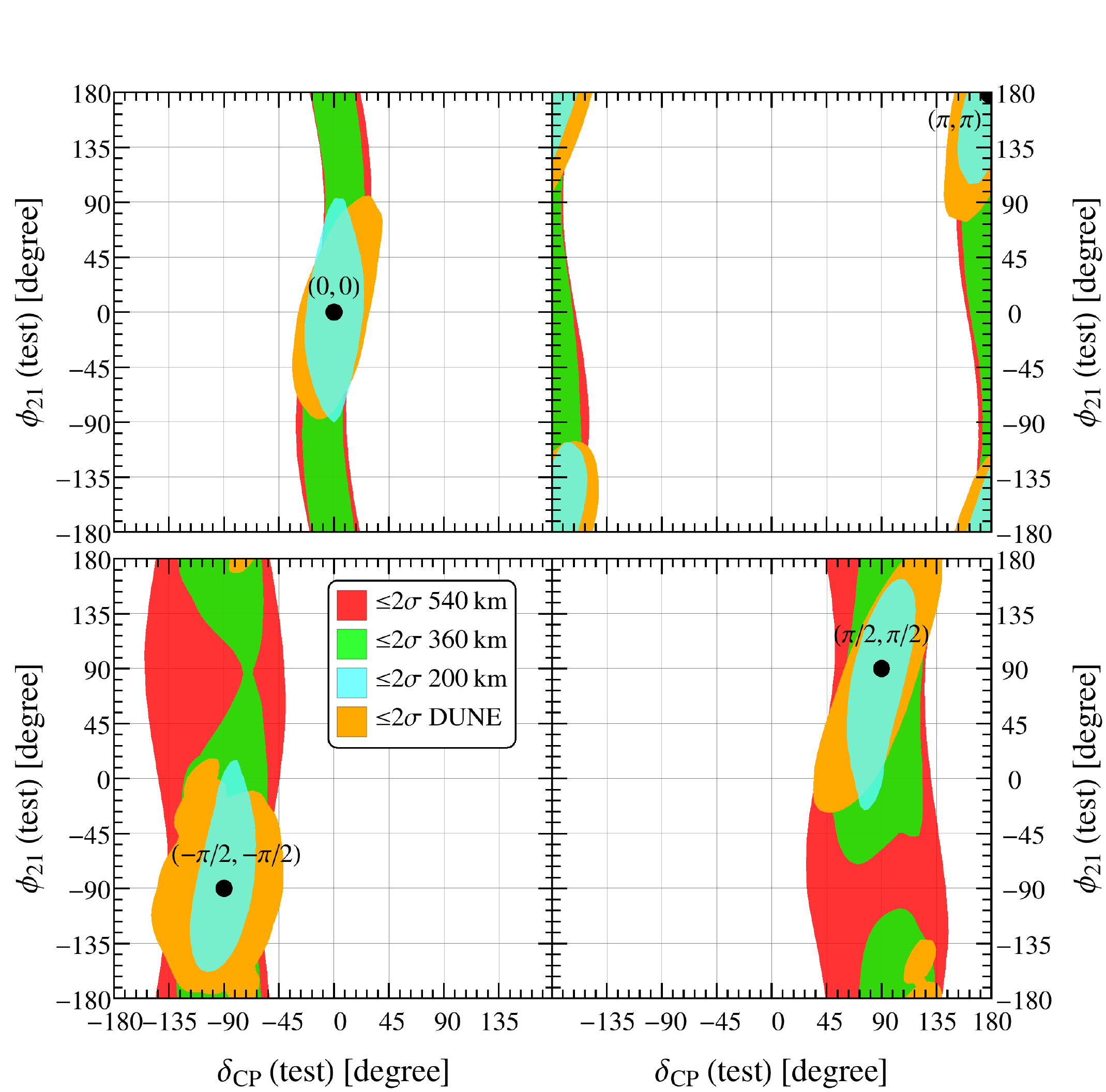}
\caption{
  CP reconstruction in the plane $\left[\delta_{\rm CP},\,\phi_{21}\right]$ (test) for different ESSnuSB baselines at $2\sigma$ level for 2 d.o.f. 
  We have fixed $|\alpha_{21}|=0.02$ and normal mass ordering in the data as well as in the theory. The red, green, and cyan contours correspond to 540, 360, 200 km baselines of the ESSnuSB experiment, respectively, whereas  the orange contours denote the sensitivity expected in DUNE.  
}
\label{cp_rec1}
\end{figure*}

  The results of our analysis are shown in Fig.~\ref{cp_rec1}.
%
%
The upper two panels correspond to the two CP conserving cases $(0,0)$ and $(\pi, \pi)$, and the lower two panels to the two CP violating scenarios $(-\pi/2, -\pi/2)$ and $(\pi/2, \pi/2)$. 
The red, green, and cyan contours in each panel correspond to 540, 360, and 200 km baselines of the ESSnuSB experiment, respectively, whereas the orange contours represent the sensitivity expected in DUNE. 
All contours correspond to the $2\sigma$ level for 2 d.o.f. 
For this analysis, we have fixed $|\alpha_{21}| = 0.02$, which lies within the current $3\sigma$ allowed boundary. 
All the other off-diagonal NU parameters have been kept fixed to zero.
We have marginalized over the mixing angles $\theta_{13}$ and $\theta_{23}$ and the atmospheric mass-squared splitting, $\Delta m_{31}^2$, as well as over the true and test values of the NU parameters $\alpha_{11}$, $\alpha_{22}$, $\alpha_{33}$ within their allowed ranges. 

We have checked that the expected $1\sigma$ uncertainties on $\delta_{\rm CP}$ ($\phi_{21}$) for the CP-violating scenarios are $16^\circ$ ($100^\circ$) for 540 km, and $13^\circ$ ($70^\circ$) for 360 km. 
 For the CP-conserving benchmarks, the uncertainties on $\delta_{\rm CP}$ are $10^\circ$ for 360 km, and $12^\circ$ for 540 km respectively, whereas no sensitivity on $\phi_{21}$ is found for these two baselines.
    On the other hand, for the 200 km baseline, the typical $1\sigma$ level uncertainty on $\delta_{\rm CP}$ ($\phi_{21}$) is $10^\circ$ ($40^\circ$).
One sees that the best performance would be obtained for the shortest baseline, 200 km.
For comparison, we have also projected the sensitivity of the DUNE experiment in the same plot.
For DUNE, the typical $1\sigma$ level uncertainty on the reconstructed CP phase $\delta_{\rm CP}$ ($\phi_{21}$) is $21^\circ$ ($45^\circ$). 

In short, one can see that the $\delta_{\rm CP}$ reconstruction capability of ESSnuSB does not get too much impaired by the presence of unitarity violation, while the sensitivity to the NU phase $\phi_{21}$ is competitive with that in DUNE and even better for 200 km baseline.
Certainly, a future combined analysis of the actual results of these experiments would improve the standard and new CP-phase reconstruction sensitivities.

\section{Summary and outlook} 
\label{sec:Conclusions}

  Here we have explored the physics potential of the proposed European Spallation Source facility in the presence of nonunitarity of the lepton mixing matrix,
  as generally expected within the seesaw paradigm. We have also explored the DUNE physics potential in this framework as well as the combined result of both
    DUNE and ESSnuSB experiments.
  First, we presented in detail the theory framework of neutrino oscillations with effective nonunitary neutrino mixing,
  discussing in Fig.~\ref{fig:prob1} the resulting neutrino and antineutrino appearance oscillation probabilities.
Throughout the paper we have assumed normal mass ordering, and considered three reference baseline choices of 540, 360, and 200 km for the ESSnuSB setup. 
In Fig.~\ref{alpha21_phi21} we have presented the sensitivity contours in the $(|\alpha_{21}|,\,  \delta_{\rm CP})$ plane. 
The promising results for the 200 km baseline were understood in terms of the expected $\nu_{\mu}\to  \nu_e$ and $\bar{\nu}_{\mu}\to  \bar{\nu}_e$ appearance event spectra.
These are given in Fig.~\ref{events_spectra} as a function of the reconstructed neutrino energy. 
We found encouraging ESSnuSB and DUNE sensitivities 
for the off-diagonal NU parameter $|\alpha_{21}|$, and the two diagonal NU parameters $\alpha_{11}$ and $\alpha_{22}$, as seen in Fig.~\ref{fig:chisq_alpha}. 
One also sees how the combined analysis of DUNE and ESSnuSB can help improving the sensitivity on the above mentioned NU parameters beyond the current $3\sigma$ bounds
\footnote{The results we have obtained are also competitive with the sensitivity expected at future coherent elastic neutrino-nucleus scattering experiments~\cite{Miranda:2020syh}.}.

More remarkable, perhaps, is the ESSnuSB CPV discovery potential, which is better than that of DUNE, as illustrated in Figs.~\ref{fig:cpv}, \ref{fig:cpv2}, and \ref{fig:cpv3}.
  One appreciates also a milder degrading of the ESSnuSB CP violation sensitivities with respect to the standard unitary mixing scenario in comparison with DUNE,
  where the effect is stronger.
ESSnuSB would therefore contribute to establishing the robustness of CP determination against small departures from unitarity arising, say, from the seesaw mechanism.
Finally, we have also obtained a promising CP reconstruction potential, both for the standard CP phase of the three-neutrino paradigm, as well as for the phase associated to nonunitarity, see Fig.~\ref{cp_rec1}.
Altogether, within the generalized nonunitary neutrino mixing framework, we have found that the proposed ESSnuSB facility is competitive and
complementary to DUNE, not only for leptonic CP violation studies, but also for probing new physics parameters associated to unitarity violation.

\vspace*{1cm}

\subsection*{ACKNOWLEDGMENTS}
\vspace*{-.3cm}
\noindent This work is supported by Spanish grants Agencia Estatal de Investigaci{\'o}n under grant no. PID2020-113775GB-I00 (AEI/10.13039/501100011033) and Prometeo CIPROM/2021/054 (Generalitat Valenciana),
in part by the European Union Horizon 2020 Research and Innovation Programme under the Marie Sklodowska-Curie grant agreement No. 860881-HIDDeN,
and by the LabEx P2IO (ANR-10-LABX-0038 - Project ``BSMNu'') in the framework of the ``Investissements d’Avenir'' (ANR-11-IDEX-0003-01 ) managed by the Agence Nationale de la Recherche (ANR), France,
and by CONACYT-Mexico under grant A1-S-23238. O. G. M. has been supported by SNI (Sistema Nacional de Investigadores, Mexico).



\end{document}